\RequirePackage[hyphens]{url}
\documentclass[journal=jpccck,manuscript=article]{achemso}

\usepackage[version=3]{mhchem} 
\usepackage{xcolor}
\usepackage{soul} 
\usepackage{lineno,hyperref}

\newcommand{\DIPC}[0]{
Donostia International Physics Center (DIPC),
Paseo Manuel de Lardizabal 4, 20018 Donostia-San Sebasti\'an, Spain}
\newcommand{\CFM}[0]{
Centro de F\'{\i}sica de Materiales CFM/MPC (CSIC-UPV/EHU), Paseo Manuel de Lardizabal 5, 20018 Donostia-San Sebasti\'an, Spain}

\author{Auguste Tetenoire}
\email{auguste.tetenoire@dipc.org}
 \affiliation{\DIPC}

\author{J.\ I\~naki\ Juaristi}
\email{josebainaki.juaristi@ehu.eus}
\affiliation{Departamento de Pol\'{\i}meros y Materiales Avanzados: F\'{\i}sica, Qu\'{\i}mica y Tecnolog\'{\i}a, Facultad de Qu\'{\i}micas (UPV/EHU), Apartado 1072, 20080 Donostia-San Sebasti\'an, Spain}
 \alsoaffiliation{\CFM}
 \alsoaffiliation{\DIPC}

\author{Maite Alducin}
\email{maite.alducin@ehu.eus}
 \affiliation{\CFM}
 \alsoaffiliation{\DIPC}

\title{Insights into the coadsorption and reactivity of O and CO on Ru(0001) and their coverage dependence}
\abbreviations{IR,NMR,UV}

\begin{document}

\begin{tocentry}

\includegraphics[width=0.75\columnwidth]{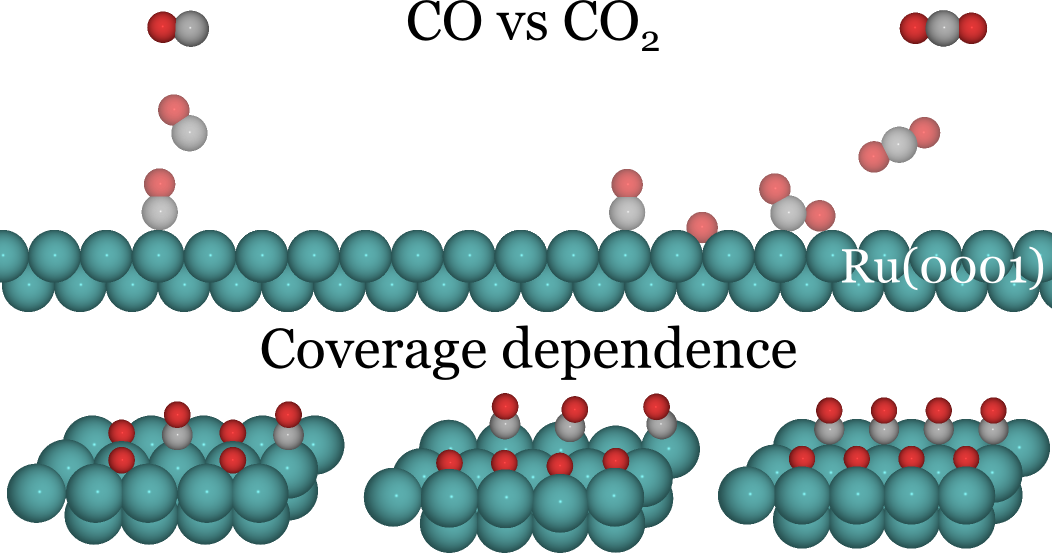}

\end{tocentry}

\begin{abstract}
Using density functional theory and an exchange-correlation functional that includes the van der Waals interaction, we study the coadsorption of CO on Ru(0001) saturated with 0.5ML of oxygen. Different coexisting CO coverages are considered that are experimentally motivated, the room temperature coverage consisting in 0.5ML-O+0.25ML-CO (low coverage), the saturation coverage achieved at low temperatures (0.5ML-O+0.375ML-CO, intermediate coverage), and the equally mixed monolayer that is stable according to our calculations but not experimentally observed yet (0.5ML-O+0.5ML-CO, high coverage). For each coverage, we study the competition between the desorption and oxidation of CO on the corresponding optimized structure by analyzing their reaction energies and minimum energy reaction paths. The desorption process is endothermic at all coverages, although the desorption energy decreases as the CO coverage increases. The process itself (and also the reverted adsorption) becomes more involved at the intermediate and high coverages because of the appearance of a physisorption well and concomitant energy barrier separating it from the chemisorbed state. Remarkably, the oxidation of CO, which is endothermic at low coverages, turns exothermic at the intermediate and high coverages. In all cases, the minimum reaction path for oxidation, that involves the chemisorbed and physisorbed CO$_2$, is ruled by one of the large energy barriers that protect these molecular states. Altogether, the larger activation energies for oxidation as compared to those for desorption and the extreme complexity of the oxidation against the desorption paths explain that CO desorption dominates over the oxidation in experiments.  
\end{abstract}

keywords: CO oxidation; CO desorption; coverage dependence; Ru(0001); heterogeneous catalysis; van der Waals density functional theory


\section{Introduction}

One of the most important heterogeneous catalytic reactions is the oxidation of CO on metal surfaces. From the technological side, this reaction is crucial for the conversion of CO into CO$_2$ in car exhaust catalytic converters, and, in general, within the context of emission control and maintenance of clean air.
From the fundamental point of view, due to its relative simplicity it has become a model system for utilizing different surface science techniques and advanced theoretical characterization methods. In particular,
great interest has been devoted to the behavior of ruthenium as a catalyst for this reaction, that has been regarded as anomalous when compared to other transition metals as palladium, platinum, rhodium, and iridium. More precisely, it was found that under ultra high vacuum (UHV) conditions Ru is very inactive for CO oxidation~\cite{lee1980,madey1975}. However, at high gas pressures Ru was found to be a much more active catalysts for CO oxidation than Pt, Pd, Rh and Ir~\cite{Peden1986kinetics,Peden1991Aug,Hoffmann1991Aug}. These results have motivated numerous experimental and theoretical studies devoted to the understanding and characterization of the adsorption and coadsorption of O$_2$ and CO and their interaction on Ru surfaces. As a consequence, a large amount of knowledge has already been gained about this system.

Regarding the dissociative adsorption of O$_2$ on Ru(0001), it has been well established using low energy electron diffraction (LEED) that under UHV conditions and at room temperature, two ordered phases can be formed, p(2$\times$2)~\cite{Lindroos1989Nov} and p(2$\times$1)~\cite{Pfnur1989Oct}, corresponding to 0.25 and 0.5 monolayer (ML) oxygen coverage, respectively. The latter corresponds to the saturation coverage with oxygen under these conditions. In both cases the oxygen atoms adsorb in hcp sites~\cite{Pfnur1989Oct,Lindroos1989Nov}. Calculations based on density functional theory (DFT) at the generalized gradient approximation (GGA) level showed that, indeed, the hcp site is the energetically most favorable adsorption site for atomic oxygen on Ru(0001) at these coverages~\cite{Stampfl1996Jul}. Interestingly, that study predicted the stability of a complete one~ML of oxygen arranged in a (1$\times$1) phase, with the oxygen atoms also adsorbed on the hcp sites. The existence of this high coverage phase was subsequently corroborated in experiments that used LEED to characterize the surface structure~\cite{Stampfl1996Oct}.
However, this coverage is not achievable upon O$_2$ exposure under UHV conditions and requires that the surface is supplied with atomic O. In ref.~\citenum{Stampfl1996Oct}, this was attained by the decomposition of NO$_2$ into adsorbed O and instantly desorbed NO. It was speculated that this phase was responsible of the above mentioned high catalytic activity for CO oxidation that Ru exhibits at high pressures~\cite{Peden1991Aug}. Additionally, DFT-GGA based studies on the energetics of the reaction at this oxygen coverage seemed to support this idea~\cite{Stampfl1997Feb,Stampfl1999Aug}. However, further CO deposition and molecular beam experiments over oxygen precovered Ru(0001) showed that the (1$\times$1)-O overlayer on Ru (0001) is very inactive, and that only for oxygen loadings beyond 3~ML the CO/CO$_2$ conversion is efficient~\cite{bottcher1997,Bottcher1999Nov,Bottcher1999jpccb}. Finally, it was shown that the active part in the oxidation of CO is an ultrathin RuO$_2$ surface oxide that grows at high O$_2$ exposures~\cite{Over2000,Over2003Jun}.

Indeed, experimental and theoretical studies on the coadsorption of O and CO on Ru(0001) have constituted also an invaluable source of information to understand the individual steps in the mechanisms of the CO oxidation in this surface. In this regard, Kostov et al.~\cite{Kostov1992Nov} using a battery of techniques such as high resolution electron energy loss spectroscopy (HREELS), LEED, temperature programmed desorption (TPD), and measurements of work function changes, performed an extensive characterization of the different structures that emerge upon the adsorption of CO on Ru(0001) with different precoverages of O. One of the important results of this work concerns the CO saturation coverage on the 0.5~ML oxygen precovered surface, which is the oxygen saturation coverage under UHV conditions. First, it was found that at 120~K CO saturation leads to crowded CO in the empty space of the (2$\times$1)-O layer. However, at 300~K around one third of the CO was desorbed so that saturation at this temperature corresponds to a O:CO ratio of roughly 2:1. Interestingly, in this high temperature structure, the experimental information strongly suggested that half of the O atoms changed place from their equilibrium hcp sites to less stable fcc sites forming a honeycomb structure, in which CO adsorbs vertically on top of the Ru atom located at the center of the oxygen hexagon. This structure had been suggested by a previous study that used Fourier transform-infrared reflection absorption spectroscopy (FT-IRAS), LEED, and thermal desorption mass spectroscopy (TDMS)~\cite{hoffmann1991situ}, and its existence was finally confirmed by a LEED-IV analysis~\cite{narloch1994}. It is remarkable that, despite the ample range of coverages analysed, formation of CO$_2$ was never observed in the thermal desorption experiments of ref.~\citenum{Kostov1992Nov}, that confirmed the low activity of Ru for CO oxidation under UHV conditions.

The theoretical studies on coadsorbed O and CO structures on Ru(0001) performed up to now have been based on DFT-GGA. In ref.~\citenum{Stampfl2002Apr} results were provided for the CO adsorption energies and adsorption activation barriers at different O and CO coverages, including adsorption phases that occur in nature and also model phases not realized in experiments yet. 
Regarding the 0.5~ML oxygen saturation coverage under UHV conditions, two different phases were studied: the above mentioned honeycomb structure with O atoms equally distributed in hcp and fcc sites and CO molecules in on-top sites, and an additional structure with O in hcp sites arranged in the p(2$\times$1) structure and the CO molecules in hcp sites. The adsorption energy of CO was found to be around 0.6~eV larger in the honeycomb structure. However, note that in both considered structures the O:CO ratio was 2:1, i.e., none of them corresponded to the low temperature CO saturation coverage (the one with crowded CO molecules in the empty space of the (2$\times$1)-O layer) found experimentally by Kostov et al.~\cite{Kostov1992Nov}. 

Also the energetics and minimum energy path for CO oxidation have been studied for different adsorbate structures and coverages, including the p(2$\times$2)-O+CO structure~\cite{Zhang1999,Zhang2000Jun, Liu2013}, the p(2$\times$1)-O with CO in hcp sites in a O:CO ratio of 2:1~\cite{Liu2013,Oberg2015Aug}, the 0.5~ML O honeycomb structure~\cite{Oberg2015Aug,Ostrom2015Feb}, and two very low coverage phases consisting in one O and one CO in a 5$\times$5 cell and three O and one CO in the same cell~\cite{Zhao2017Jun}. A general result of these studies was that the main responsible of the activation barrier that hinders CO oxidation at Ru(0001) in UHV conditions is the energy required to destabilize and move the atomic oxygen from its adsorption site. In most cases it was also found that the adsorption energy of CO was larger than the corresponding activation energy for CO oxidation, with the exception of the very low coverage results of ref.~\citenum{Zhao2017Jun}.

Although Kostov et al. showed that CO oxidation cannot be thermally activated under UHV conditions~\cite{Kostov1992Nov}, it has been demonstrated that the reaction can be efficiently propelled by exciting the system with electromagnetic radiation~\cite{Bonn1999Aug,Oberg2015Aug,Ostrom2015Feb,Larue2015Jul,Schreck2018Dec}.  For instance, both CO and CO$_2$ are desorbed when the surface is excited using ultraviolet, visible, and near-infrared femtosecond laser pulses~\cite{Bonn1999Aug,Oberg2015Aug,Ostrom2015Feb}. A proper characterization of this kind of experiments requires simulations of the adsorbates dynamics in a highly excited system accurately describing the adsorbate-surface interaction at the DFT level~\cite{vazhappilly2009,fuchsel10,fuchsel11,loncaricprb16,Scholz2016Oct}. In particular, the newly developed ab initio molecular dynamics with electronic friction model~\cite{novkoprb15,novkoprb16,novkoprb17}, and its extension to incorporate the effect of a heated electronic system~\cite{juaristiprb17,alducinprl19,Scholz19}, constitutes a promising tool to the study of photoinduced CO oxidation at surfaces.

Before performing the dynamics simulations, the first step consists in the characterization of the initial state of the system prior to its excitation with the laser pulse. In the experiments of refs.~\citenum{Bonn1999Aug,Oberg2015Aug,Ostrom2015Feb} the Ru(0001) surface is first dosed with oxygen up to saturation, which in the chosen conditions it means that a 0.5~ML O coverage is achieved~\cite{Kostov1992Nov}. Subsequently, the system is dosed with CO up to saturation. In the theoretical calculations of refs.~\citenum{Oberg2015Aug,Ostrom2015Feb} it was assumed that this corresponded to a 0.25~ML CO coverage, arranged in the honeycomb structure~\cite{Oberg2015Aug,Ostrom2015Feb} or in a p(2$\times$1)-O arrangement with the CO in near atop sites~\cite{Oberg2015Aug}. However, after surface preparation and before exciting it with the laser pulse, in these experiments the surface is cooled down to 100~K. According to Kostov et al.~\cite{Kostov1992Nov}, at this temperature the equilibrium structure consists in a p(2$\times$1)-O arrangement with crowded CO at the empty space of the layer. In other words, it cannot be excluded that under the experimental conditions of refs.~\citenum{Bonn1999Aug,Oberg2015Aug,Ostrom2015Feb} the CO coverage is higher than 0.25~ML. Therefore, it would be interesting to perform dynamics simulations, as those discussed above, for different coverages of CO and adsorbate structures, provided they can be considered realistic in the light of the reported surface preparation and the observations of Kostov et al.~\cite{Kostov1992Nov}.

Motivated by these facts, here we proceed to undertake the first step of this program, which consists in performing a complete DFT characterization of different (O, CO) mixed coverages on Ru(0001) that consist of a fixed 0.5~ML of O combined with three different CO coverages, which will be denoted as low, intermediate, and high coverages. Specifically, the low coverage is formed by 0.5~ML O + 0.25~ML CO and it represents the CO saturation coverage found in ref.~\citenum{Kostov1992Nov} at 300~K. The intermediate coverage, consisting of 0.5~ML O + 0.375~ML CO, corresponds closely to the CO saturation coverage reported in ref.~\citenum{Kostov1992Nov} at temperatures below 120~K. Finally, we also study the high coverage defined by 0.5~ML O + 0.5~ML CO, despite the fact that this mixed ML has not been experimentally observed yet. Here, it will be shown that this high coverage is indeed energetically stable, but unreachable by the sample preparation procedure followed in refs.~\citenum{Kostov1992Nov,Bonn1999Aug,Oberg2015Aug,Ostrom2015Feb}.

The paper is organized as follows. The theoretical methods and computational settings used to model the different (O,CO)-covered Ru(0001) surfaces are described in the Methods section. Next, we provide and discuss the results obtained for each of the three coverages considered. In each case, we start with a systematic search of the energetically most stable configuration that is compatible with the surface preparation reported in refs.~\citenum{Kostov1992Nov,Bonn1999Aug,Oberg2015Aug,Ostrom2015Feb}. Once the optimized arrangement is identified, we perform a full characterization of the desorption and oxidation of CO on this specific overlayer by calculating the corresponding reaction energies and, importantly, the minimum energy reaction paths for each process. The comparative analysis of their energy diagrams is relevant to understand the competition between CO desorption and oxidation that is observed in experiments. 
The dependence of this competitive reactions on the three coverages considered here is discussed at the end of the Results section. The main conclusions are summarized in the Summary section.

\section{Methods}

All calculations are performed with the Vienna ab initio simulation package (VASP)~\cite{vasp1,vasp2} using density functional theory (DFT) and the van der Waals exchange-correlation functional (vdW-DF) proposed by Dion et al.~\cite{Dion2004Jun} (see the Supporting Information for details on this functional choice). Since the spin of the open-shell O atoms is quenched when adsorbed on Ru(0001)~\cite{Liem2004Sep,Legare2005Apr,Xin2015Apr}, all the calculations are based on non spin-polarized DFT. For each atomic configuration, the electronic ground state is determined by minimizing the system total energy up to a precision of $10^{-6}$~eV. In this process, integration in the Brillouin zone is performed using $\Gamma$-centered Monkhorst-Pack grids of special \textbf{k} points~\cite{monkhorst76} (11$\times$11$\times$11 for bulk Ru and 3$\times$6$\times$1 for pristine and (O,CO)-covered Ru(0001)) and the Methfessel and Paxton scheme of first order with a broadening of 0.1~eV to describe partial occupancies of each state~\cite{mepaprb89}. The latter are expanded in a plane-wave basis set with an energy cut-off of $400$~eV, whereas the electron-core interaction is treated with the projected augmented wave (PAW) method~\cite{Blochl1994Dec} that is implemented in VASP~\cite{pawvasp}. 

Using these computational settings, the calculated bulk Ru lattice parameters ($a=2.75$~{\AA} and $c=4.32$~{\AA}) are in good agreement with  the experimental values ($a=2.70$~{\AA}, $c=4.28$~{\AA})~\cite{ashcroft1976} and with previous theoretical calculations~\cite{Zhao2017Jun} ($a=2.74$~{\AA}, $c=4.33$~{\AA}). Next, the pristine and the (O,CO)-covered Ru(0001) surfaces are both modeled using the same supercell, that is, a periodic five-layer slab defined by a (4$\times$2) surface unit cell and $19$~{\AA} of vacuum. This large supercell allows us to investigate the three (O,CO) coverages of interest, while assuring negligible spurious interactions between periodic images. The 0.5~ML of oxygen common to the three mixed coverages is described by four adsorbed O atoms, whereas two, three, and four CO molecules are respectively used to model the low, intermediate, and high coverages. 
The relaxed Ru(0001) structure is obtained by minimizing all the atomic forces in the three first layers below $0.01$~eV/{\AA}, while the two bottom layers are kept frozen. For the relaxed covered surfaces, all forces in the adsorbate adlayer are also minimized. The stability of the optimized structures is further confirmed by a normal mode calculation of the (O,CO) overlayer. The Hessian matrix is calculated with VASP using central finite differences. In the SI, we provide the frequencies of the in-phase and out-of-phase C--O internal stretch modes, which are the usually relevant ones in experiments.

The CO desorption energy is calculated as,
\begin{equation}
E=E_{\mathrm{FS}}-E_{\mathrm{IS}} \, ,  
\label{eq:E}
\end{equation}
where $E_{\mathrm{IS}}$ is the energy of the relaxed (O,CO)-covered surface under study and $E_{\mathrm{FS}}$ is the energy of the new relaxed system in which one CO is removed from the covered surface and located halfway the vacuum region. Thus, positive (negative) values of $E$ correspond to an endothermic (exothermic) desorption process. Conversely, the reverted adsorption process will be endothermic (exothermic) for $E<0$ ($E>0$) if we use eq.~\eqref{eq:E}. Note that irrespective of the coverage considered, the equilibrium bond length of the desorbed CO is 1.145~{\AA}, to be compared to the experimental value of 1.128~{\AA}~\cite{handbook}. Let us remark at this point that prior starting the coverage study, we first verified the adequacy of the vdw-DF exchange-correlation functional and our computational settings by calculating the desorption energy of CO from the pristine Ru(0001) surface. Our value of 1.666~eV is in excellent agreement with the reported experimental value of $1.658$~eV~\cite{pfnur1983influence} and previous DFT calculations that include van der Waals corrections~\cite{fuchsel14,Scholz2016Oct,Loncaric2017Oct}.

The CO oxidation energy on the surface, defined as the recombinative desorption of one adsorbed O and one adsorbed CO, is similarly calculated with eq.~\eqref{eq:E}. The initial state and hence its energy $E_{\mathrm{IS}}$ are the same as before, but the final state is now the relaxed system in which one O and one CO are removed from the covered surface and located halfway the vacuum region forming the relaxed CO$_2$ molecule. Regardless of the coverage, the latter adopts the expected gas-phase linear configuration  with a theoretical C--O distance of 1.179~{\AA} to be compared to the experimental C--O bond length of 1.160~{\AA}~\cite{handbook}.

In order to characterize the oxidation of CO on the covered Ru(0001) surfaces, CO$_\mathrm{(ads)}+$ O$_\mathrm{(ads)}$ $\rightharpoonup$ CO$_{2\mathrm{(gas)}}$, the first step is to identify possible intermediate adsorption states along the reaction path (see the Supporting Information for details on the systematic search we followed). For all coverages, we find two stable intermediate states, a chemisorbed bent CO$_2$ (bCO$_2$) and a physisorbed linear CO$_2$ (lCO$_2$). A charge state analysis is performed to determine the chemisorption or physisorption nature of these states (see below). Therefore, the oxidation of CO on each covered surface is assumed to proceed through these two CO$_2$ adsorption states as follows: (i) CO$_\mathrm{(ads)}$ + O$_\mathrm{(ads)}$ $\rightharpoonup$ bCO$_2$, (ii) bCO$_2$ $\rightharpoonup$ lCO$_2$, and (iii) lCO$_2$ $\rightharpoonup$ CO$_\mathrm{2(gas)}$. The characterization of each reaction subpath, including identification of eventual transition states (TS), is next performed by means of the climbing image nudged elastic band (CI-NEB) method~\cite{Henkelman2000Dec} using four images. 
Altogether the complete minimum energy path (MEP) for CO oxidation at each coverage is characterized by 14 intermediate states. Let us remark that zero point energy corrections have been neglected in the energetics studies for CO oxidation and desorption discussed in next section. Their effect in the reaction and activation energies is very minor (see section S6 in the Supporting Information).

Finally, the charge state of each adsorbate is estimated at selected steps of the CO oxidation MEP as, 
\begin{equation}
Q=Z-Q_{\mathrm{BC}} \, ,
\label{eq:Q}
\end{equation}
where $Z$ and $Q_{\mathrm{BC}}$ are the atomic number and Bader charge~\cite{bader} of the considered adsorbate. In the case of molecular species, both $Z$ and $Q_{\mathrm{BC}}$ are summed for all the atoms forming the molecule. In our case, the Bader charge is calculated with the implementation by Tang et al.~\cite{Tang2009} and Henkelman et al.~\cite{HENKELMAN06}. Recall that in using eq.~\eqref{eq:Q}, negative (positive) values of $Q$ indicate that the adsorbate has captured (lost) $|Q|$ electrons.

As a final remark, we acknowledge that the images of the system structures, depicted in some of the figures, have been done using the ASE python's package~\cite{ase-paper}.

\section{Results and Discussion}

\subsection{Low coverage: 0.5~ML O + 0.25~ML CO}
\begin{figure}
  \includegraphics[width=0.75\columnwidth]{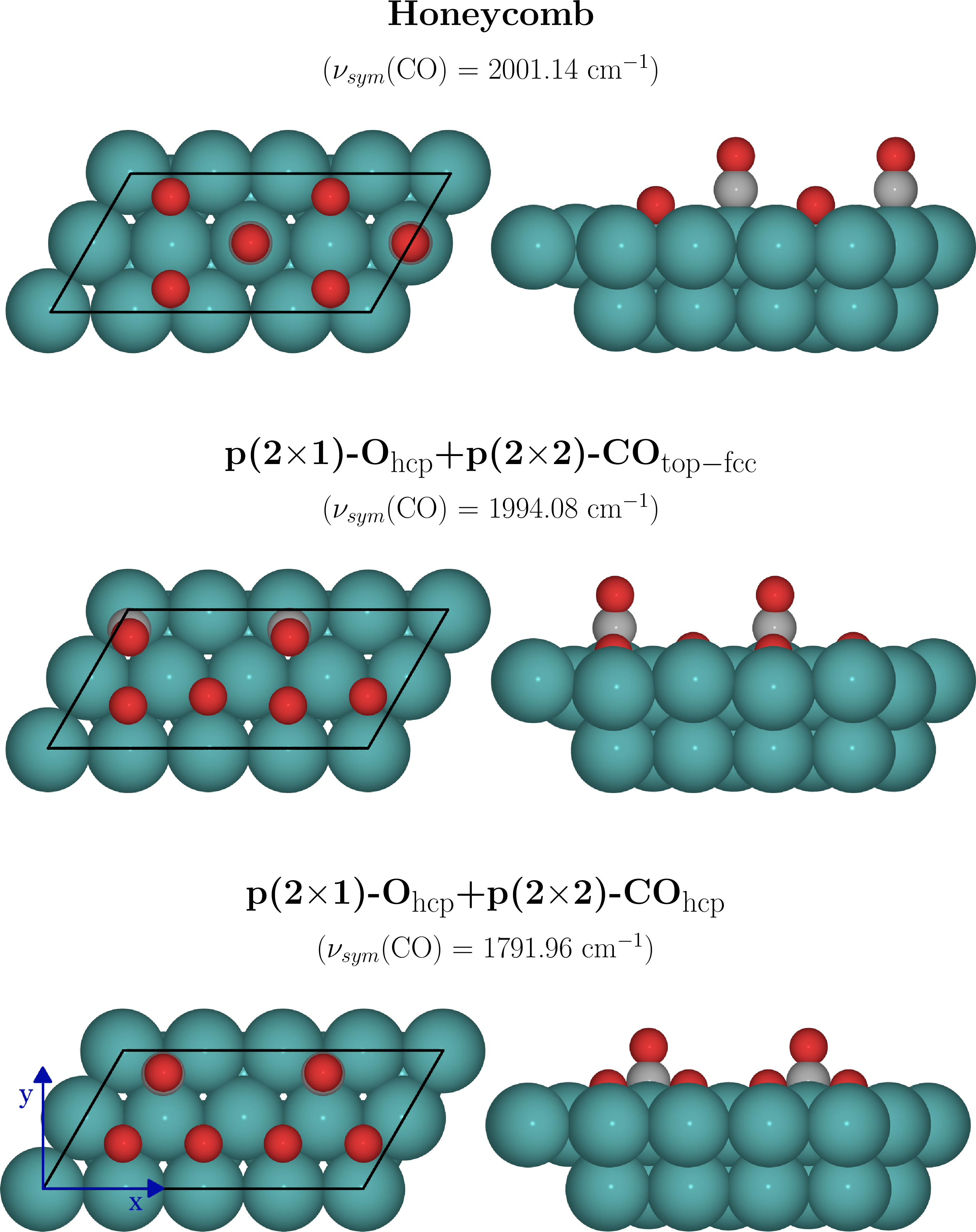}
  \caption{Top (left) and side (right) views of the three stable structures obtained for the 0.5~ML O + 0.25~ML CO coverage on Ru(0001). Top panels: the energetically most stable honeycomb structure. Middle panels: p(2$\times$1)-O$_\mathrm{hcp}$+p(2$\times$2)-CO$_\mathrm{top-fcc}$. Bottom: p(2$\times$1)-O$_\mathrm{hcp}$+p(2$\times$2)-CO$_\mathrm{hcp}$. The frequency of the in-phase C--O stretch mode is specified for each coverage. Color code: O atoms in red, C in gray, and Ru in blue. The black parallelograms show the surface unit cells in the calculations. For clarity, the periodic images of the O and CO adsorbates are not shown and only the two topmost Ru layers are depicted in the side views.}
   \label{fig:075ml_relaxation}
\end{figure}

The honeycomb structure has been confirmed experimentally as the most likely adsorbate arrangement at this coverage~\cite{hoffmann1991situ,Kostov1992Nov,narloch1994}. In this structure, each CO adsorbs atop a Ru atom and is surrounded by six oxygen atoms that occupy the second nearest hcp and fcc sites forming a honeycomb arrangement (see Fig.~\ref{fig:075ml_relaxation}, top panel). To confirm whether the honeycomb structure is indeed the energetically most favorable adsorbate arrangement at this coverage, we additionally study the stability of other possible structures that are compatible with the initial oxygen-saturation coverage of 0.5~ML that is formed in the fs-laser induced reaction experiments prior dosing the surface with CO~\cite{Oberg2015Aug,Ostrom2015Feb,Bonn1999Aug}. In this respect, it is well established by experiments and theory that at 0.5~ML coverage and in absence of CO adsorbates, the O atoms adsorb preferentially in hcp sites forming a p(2$\times$1) structure~\cite{Pfnur1989Oct,Stampfl1996Jul}. Therefore, coadsorbed with p(2$\times$1)-O$_\mathrm{hcp}$, we consider four possible p(2$\times$2)-CO structures that correspond to the two CO molecules in our 4$\times$2 unit cell being initially at either top, bridge, fcc, or hcp sites.
After relaxation, the CO molecules change their initial positions in the three first cases and, at the end, only the two optimized structures depicted in Fig.~\ref{fig:075ml_relaxation} for the p(2$\times$1)-O$_\mathrm{hcp}$ arrangement are stable. The stability is further confirmed by the absence of imaginary normal modes in the three optimized (O, CO) overlayers. Table~\ref{tab:075ml_adsorption} summarizes all the results of the structural search performed at this low coverage. The frequencies of the in-phase C--O internal stretch mode (i.e., when the two CO vibrate in phase) are written in Fig.~\ref{fig:075ml_relaxation} for each optimized structure. The lowest value corresponds to the p(2$\times$2)-CO$_\mathrm{hcp}$ arrangement, for which the CO molecules locate closer to the surface. The frequencies for the honeycomb and p(2$\times$2)-CO$_\mathrm{top-fcc}$ only differ in 7~cm\textsuperscript{-1}, showing that the CO adsorption properties are locally similar in both cases. 

In agreement with experiments, the honeycomb structure is energetically the most stable one. In the case of the p(2$\times$1) oxygen arrangement, coadsorption of CO on the top site, which is the preferred adsorption site in the honeycomb structure~\cite{hoffmann1991situ,Kostov1992Nov,narloch1994} and in the zero coverage limit~\cite{michalk83,Stampfl2002Apr,Scholz2016Oct,Loncaric2017Oct}, is unstable since all the CO desorb upon relaxation. Optimization of the two structures in which the CO molecules are initially located on either bridge or fcc sites, leads in both cases to the same p(2$\times$2)-CO$_\mathrm{top-fcc}$ structure depicted in Fig.~\ref{fig:075ml_relaxation}. In this relaxed overlayer, the CO molecules end up with the center of mass in the line joining the top and fcc sites, with their axis slightly tilted towards the nearest O atom. 
Additionally, the farthest O atoms shift about $0.32$~{\AA} along the $y$ direction. 
This is the most stable structure with the p(2$\times$1) oxygen arrangement, but it is still $0.759$~eV higher in energy per simulation cell than the honeycomb structure. Finally, relaxation of the structure with the CO molecules in the unoccupied hcp sites shows that this arrangement is also stable, but 0.359~eV higher in energy per simulation cell than the previous one. In this respect, it is worth to mention that previous DFT studies of the same coverage with the p(2$\times$1) oxygen arrangement had considered different adsorption positions for the CO molecules: the hcp sites in ref.~\citenum{Liu2013} and near atop sites in ref.~\citenum{Oberg2015Aug}. Our DFT+vdW-DF results show that for this oxygen arrangement the two proposed structures are stable. Nevertheless, they also suggest that the top-fcc location may be a more realistic adsorption site for the coadsorbed CO, in agreement with ref.~\citenum{Oberg2015Aug}.
Next, the analysis concerning the energetics of CO desorption and oxidation on the low coverage will be performed on the honeycomb structure, which is confirmed as the lowest energy configuration.

\begin{table}
    \centering
    \begin{tabular}{cccccc}
       \hline\hline
        O-structure & O site & CO initial site & CO final site & $E$(eV) & $Z_{\mathrm{CO-Ru}}$({\AA})\\
       \hline
        Honeycomb       & 0.5fcc+0.5hcp & top       & top       & 0.000    & 2.63 \\
        p(2$\times$1)       & hcp           & top    & desorbed  & --    & -- \\
        p(2$\times$1)       & hcp           & bridge    & top-fcc  & 0.759    & 2.67 \\
        p(2$\times$1)   & hcp             & fcc       & top-fcc  & 0.759    & 2.67 \\
        p(2$\times$1)               & hcp             & hcp       & hcp       & 1.118    & 2.16 \\
        \hline\hline
    \end{tabular}
  \caption{Results from the structural optimization for the 0.5~ML O + 0.25~ML CO coverage on Ru(0001), indicating the structure and adsorption sites of the O atoms, the initial and final adsorption sites of the coadsorbed (2$\times$2)-CO, the potential energy per simulation cell $E$ referred to that of the lowest energy honeycomb structure, and the height of the CO center of mass from the surface (defined as the average heights of the Ru atoms in the topmost layer) $Z_{\mathrm{CO-Ru}}$.\label{tab:075ml_adsorption}}
\end{table}

\subsubsection{CO desorption and oxidation on the honeycomb structure}

\begin{table}
    \centering
    \begin{tabular}{c c c}
        \hline\hline
        Reaction  & Adsorption site         & $E$ (eV)\\
        \hline
        CO$_\mathrm{(ads)} \rightharpoonup $ CO$_\mathrm{(gas)}$          & CO$_\mathrm{top}$           & 1.569 \\
& & \\
        O$_\mathrm{(ads)}$+CO$_\mathrm{(ads)} \rightharpoonup $ CO$_\mathrm{2(gas)}$      & CO$_\mathrm {top}$, O$_\mathrm {fcc}$    & 0.643 \\
         &    CO$_\mathrm {top}$, O$_\mathrm {hcp}$     & 1.206 \\
    \hline\hline
    \end{tabular}
   \caption{Reaction energies $E$ for CO desorption and oxidation from the Ru(0001) surface with 0.5~ML O + 0.25~ML CO coverage in the honeycomb structure. The initial adsorption site for each desorbing species is indicated as a subscript in the second column.}
    \label{tab:075ml_desorption}
\end{table}

The reaction energies for CO desorption and oxidation on the honeycomb structure are summarized in Table~\ref{tab:075ml_desorption}. The CO desorption energy of 1.569~eV, which is around 0.1~eV less endothermic than on the pristine surface, would be in line with existing DFT calculations that report a decrease in the CO desorption energy as the O coverage increases~\cite{Stampfl2002Apr,Liu2013}. Nevertheless, the values we obtain in both cases are rather similar, suggesting that the CO desorption energetics is not much influenced by the presence of O adsorbates. Oxidation on the honeycomb structure is also endothermic irrespective of whether the CO recombines with the O adsorbed at either the fcc site (0.643~eV) or the hcp site (1.206~eV). The lower oxidation energy obtained in the former case is consistent with the higher binding energy of O in the hcp site than in the fcc site~\cite{Stampfl1996Jul,shimizu2008,Cai2015Dec,Zhao2017Jun}. Interestingly, the oxidation energy is in both cases smaller than the desorption energy. Nonetheless, this result alone is insufficient to determine the likeliness of one process over the other, since it may hide the existence of energy barriers along the reaction paths. This is precisely the kind of information provided by the MEPs that we analyze next to further characterize the competition between both processes.

\begin{figure}
  \includegraphics[width=0.75\columnwidth]{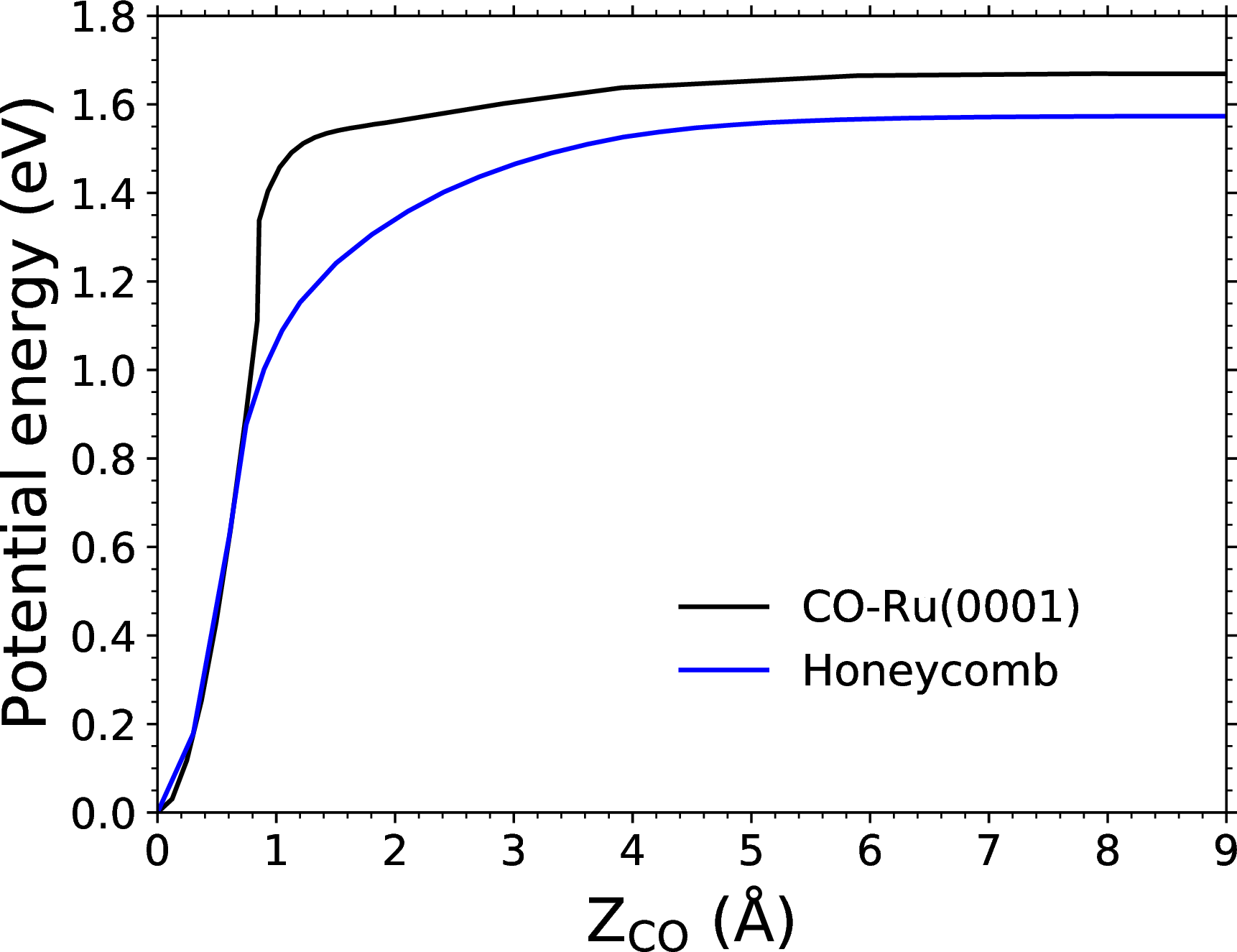}
 \caption{Potential energy of a CO molecule as a function of its center of mass height $Z_{\mathrm{CO}}$, measured from its equilibrium adsorption position $Z_{\mathrm{CO-Ru}}$. Results obtained for the honeycomb structure (blue) and for the pristine Ru(0001) surface (black). The zero of energy is chosen as that of the equilibrium adsorption position for each coverage.}
  \label{fig:075ml_co_desorption}
\end{figure}

Figure~\ref{fig:075ml_co_desorption} shows the potential energy against the CO center of mass (CM) height $Z_{\mathrm{CO}}$ that is measured from its adsorption position in the honeycomb structure. For comparison, we also show the potential energy curve for CO desorption in the case of the pristine Ru(0001) surface (denoted as CO-Ru(0001) hereafter). In each curve, the equilibrium configuration with the molecule adsorbed in its adsorption well atop a Ru atom is taken as the energy reference. At distances $Z_{\mathrm{CO}}>6$~{\AA} the curves have already reached the asymptotic region and the energy values basically coincide with the desorption energies. Both curves have as common important features the apparent absence of a physisorption well and hence the absence of energy barriers for CO adsorption. The latter is important regarding experiments in which the coadsorption of O and CO on surfaces is realized by first adsorbing O up to the required coverage and next adsorbing CO under thermal deposition. This is the procedure used, for instance, in refs.~\citenum{Kostov1992Nov,Bonn1999Aug,Oberg2015Aug,Ostrom2015Feb}. The absence of energy barriers for CO adsorption means that the 0.5~ML O + 0.25~ML CO honeycomb structure is achievable using this procedure. We note in passing that the CO-Ru(0001) desorption curve is remarkably similar to the one calculated in ref.~\citenum{Scholz2016Oct} with RPBE-D2, using a (2$\times$2) surface cell and different computational settings.

The energetics of CO oxidation on the honeycomb structure is more involved. For simplicity, we will focus on the reaction between the CO and the O adsorbed in the fcc site, which is the energetically most favorable recombination (see Table~\ref{tab:075ml_desorption}). As above mentioned, the oxidation is assumed to proceed through the two molecular adsorption states that are identified in this system, the chemisorbed bCO$_{2}$ and the physisorbed lCO$_{2}$. It is worth to remark that intermediate states of similar characteristics have been obtained in the reaction path for CO oxidation on Pt(111)~\cite{Zhou2019Mar,Wu2019Apr} and Co(0001)~\cite{Liu2016}. On Ru(0001), Zhao et al.~\cite{Zhao2017Jun} at low CO and O coverage and \"Ostr\"om et al.~\cite{Ostrom2015Feb} for the same honeycomb structure analysed here, described also the bCO$_2$ but not the lCO$_2$ as intermediate state in the MEP for CO oxidation. This is probably due to the fact that the DFT calculations in both references were based on GGA exchange-correlation functionals. Such functionals, not including van der Waals corrections, are expected to be unable to describe the physisorption region relevant for the characterization of the lCO$_2$ state.

\begin{figure}
  \includegraphics[width=0.75\columnwidth]{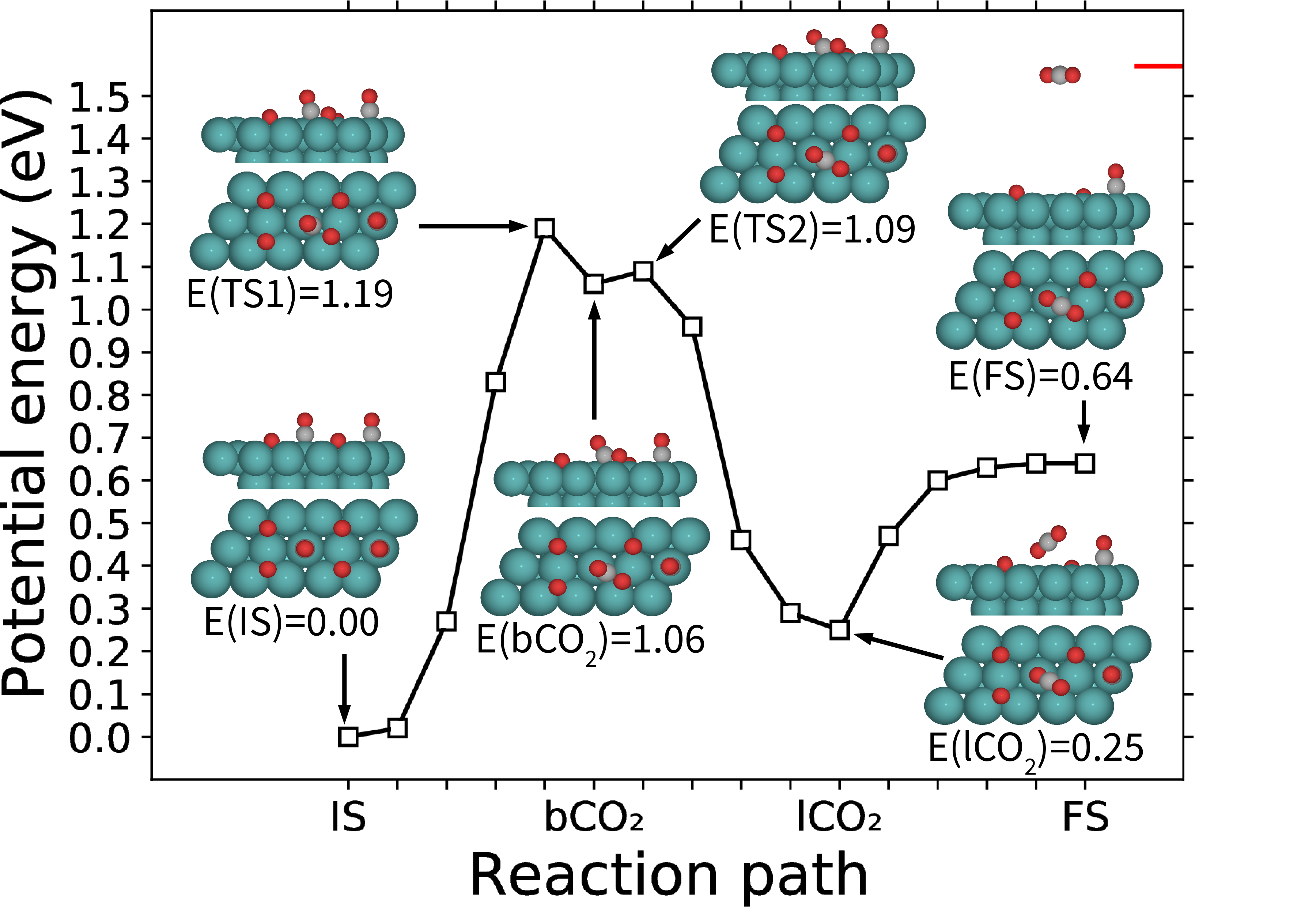}
  \caption{CI-NEB calculated minimum energy path for the CO$_2$ recombinative desorption from Ru(0001)-(0.5~ML O + 0.25~ML CO) in the honeycomb structure. Also are shown the top and side views of the relevant configurations along the path: the initial state O$_\mathrm{(ads)}$+CO$_\mathrm{(ads)}$ (IS), the final state CO$_2\mathrm{(gas)}$ (FS), the intermediate adsorption states (bCO$_2$ and lCO$_2$), and the transition states (TS1 and TS2). Their energies referred to IS are given in eV below each image. For comparison, the energy of the desorbed CO is also given by a red short-line on the top right of the figure. Color code: O in red, C in gray, and Ru in blue.}
  \label{fig:075ml_cineb}
\end{figure}

Figure~\ref{fig:075ml_cineb} shows the MEP for the oxidation of CO with the O adsorbed in the fcc site. A detailed description of all the calculated states along the MEP are given in the Supporting Information. Here the figure shows schematically images of the initial (IS) and final (FS) states, the two molecular adsorption states (bCO$_2$ and lCO$_2$), and the two transition states we find (TS1 and TS2). Specifically, TS1 is located in the path between IS and bCO$_2$, whereas TS2 appears in the path between bCO$_2$ and lCO$_2$. It has been verified that TS1 and TS2 are transition states by checking that the frequency of the normal mode along the reaction coordinate is imaginary.

The minimum energy configuration along the CO oxidation MEP is IS. The chemisorbed bCO$_2$ is a metastable adsorption state. The molecules in this state only require 0.03~eV to desorb through the more stable lCO$_2$ state and 0.13~eV to dissociate on the surface as O$\mathrm{(ads)}$+CO$\mathrm{(ads)}$. In contrast, the physisorbed lCO$_2$ is rather stable. Energy barriers of 0.39 and 0.94~eV separate this state from desorbing and from dissociating on the surface, respectively. Regarding the oxidation reaction of interest, the overall process is governed by TS1 and it requires an activation energy of 1.19~eV. Note however that this activation energy is still 0.38~eV lower than the energy required to desorb CO (see Table~\ref{tab:075ml_desorption}). Clearly, the dominant desorption over oxidation reported in experiments has to be explained in terms of the complex oxidation process. CO$_2$ desorption requires destabilization of a strongly bound O atom, diffusion of CO and O at the surface, and the encounter between CO and O under proper geometrical and energy conditions in order to form the molecular bond. As a consequence, among all the configurations of the system that can be explored in a given dynamics, a relatively small number of them are expected to lead to oxidation, i.e., the configurational space for oxidation is much reduced as compared to the direct CO desorption. In this respect, recent experimental and theoretical studies have shown the complexity of O diffusion on a crowded CO Ru(0001) surface~\cite{Henss2019Feb,sakong20}. The way the energy is provided to the system by exciting directly the adsorbates or indirectly by thermal activation or generating electronic excitations, is also determinant regarding the relative probability for CO and CO$_2$ desorption. A final answer to these questions requires to go beyond the scope of the static analysis of the present paper and to perform a dynamical study of the relevant reaction processes. 

\subsection{Intermediate coverage: 0.5~ML O + 0.375~ML CO}

\begin{figure}
  \includegraphics[width=0.75\columnwidth]{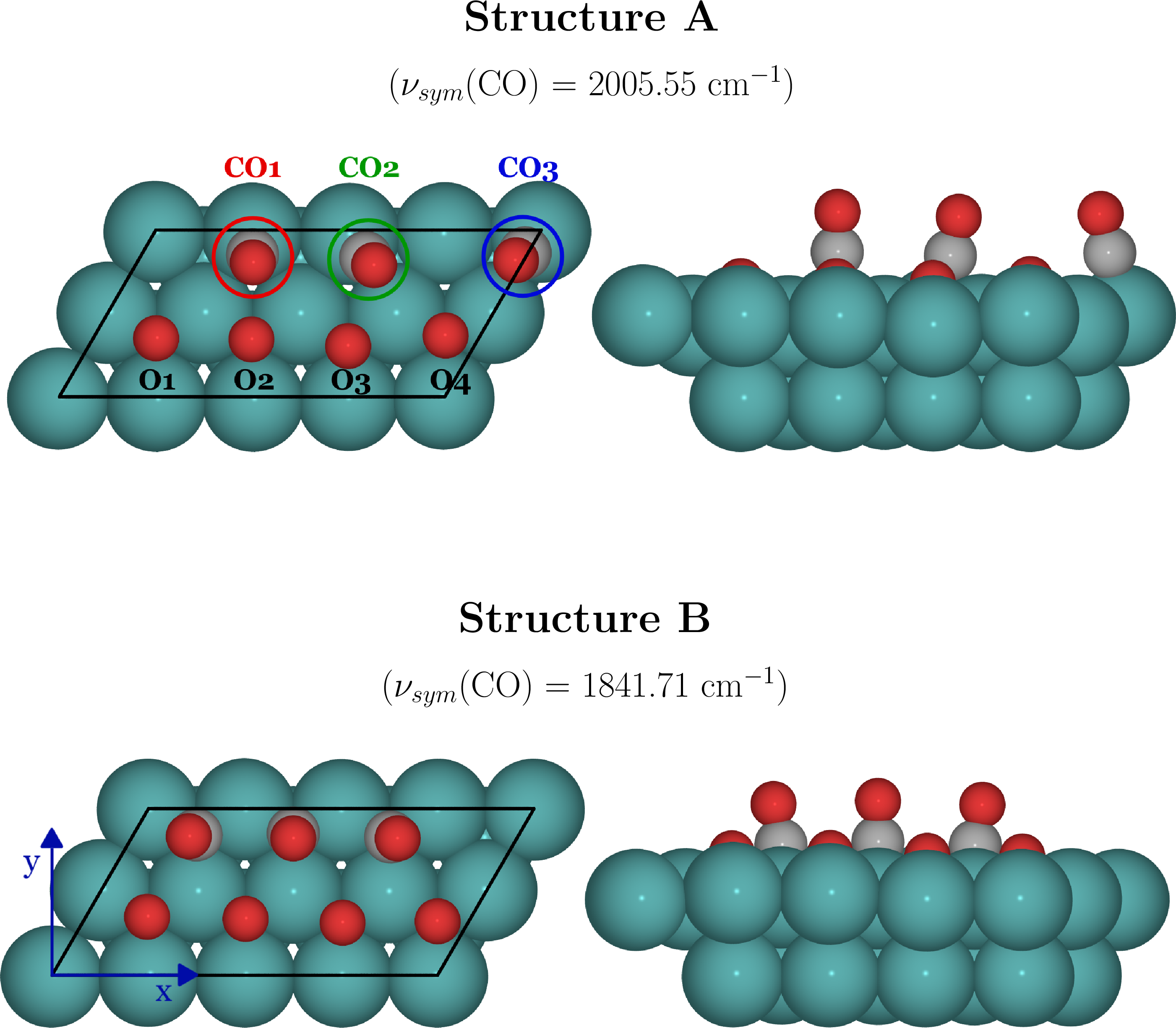}
\caption{Top (left) and side (right) views of the two stable structures obtained for the 0.5~ML O + 0.375~ML CO coverage on Ru(0001). Top panels: Structure A, which is energetically the most stable. Bottom panels: Structure B. The frequency of the in-phase C--O stretch mode is specified for each coverage. Color code: O atoms in red, C in gray, and Ru in blue. The black parallelograms show the surface unit cells in the calculations. For clarity, the periodic images of the O and CO adsorbates are not shown and only the two topmost Ru layers are depicted in the side views. The nomenclature used in the text to denote the different adsorbates in structure A is indicated in the top view. From left to right the adsorbed CO molecules are labeled as CO1, CO2, and CO3 and the adsorbed O atoms as O1, O2, O3, and O4.}
  \label{fig:0875ml_relaxation}
\end{figure}

The intermediate coverage, which is simulated by four adsorbed O and three adsorbed CO molecules in our 4$\times$2 simulation cell, is very close to the saturation coverage that was identified in experiments performed under UHV and temperatures below 120~K~\cite{Kostov1992Nov}. In this coverage, the O atoms adsorb at hcp sites in a p(2$\times$1) arrangement and the CO molecules occupy the empty space left between the O arrays. However, the precise sites in which the crowded CO are located is experimentally undetermined. Therefore, as a first step, we look for the energetically most favorable CO arrangement that coadsorbs with the p(2$\times$1)-O overlayer. Four different structures are considered in which the three CO are initially adsorbed at either top, fcc, bridge, or hcp positions. After relaxation, only the two final structures shown in Fig.~\ref{fig:0875ml_relaxation} are obtained. Both are stable structures according to the normal mode analysis. The frequencies of the in-phase internal stretch mode of the CO overlayer are written in the figure for each structure. As expected, the frequency is larger in structure A than in structure B, i.e., for the CO molecules adsorbed on near-top than in hcp sites. The frequency of 2005.55~cm\textsuperscript{-1} found for structure A compares well (within the expected DFT accuracy) with the intense HREELS peaks at 2089 and 2060~cm\textsuperscript{-1} reported for this coverage in ref.~\citenum{thomas79} and \citenum{Kostov1992Nov}, respectively. Furthermore, our calculations suggest that the low intensity peak at 1850~cm\textsuperscript{-1} that is additionally identified in ref.~\citenum{Kostov1992Nov}, but not in ref.~\citenum{thomas79}, might be related to the presence of very minor domains with, for instance, the structure B arrangement.

\begin{table}
    \centering
    \begin{tabular}{c c c c c c}
        \hline\hline
        O-structure & O site & CO initial site & CO final site & $E$(eV) & $Z_{\mathrm{CO-Ru}}$({\AA})\\
        \hline
        p(2$\times$1)      & hcp           & fcc       & near-top$^{\ref{tab:0875ml_adsorption}.A}$  & 0.000     & 2.61 \\
        p(2$\times$1)              & hcp             & bridge    & near-top$^{\ref{tab:0875ml_adsorption}.A}$  & 0.000     & 2.61
    \\
        p(2$\times$1)               & hcp             & top       & near-top$^{\ref{tab:0875ml_adsorption}.A}$  & 0.000     & 2.61 \\
        p(2$\times$1)              & hcp             & hcp       & hcp$^{\ref{tab:0875ml_adsorption}.B}$       & 0.251     & 2.20 \\
        \hline\hline
    \end{tabular}
  \caption{Results from the structural optimization for the 0.5~ML O + 0.375~ML CO coverage on Ru(0001), indicating the structure and adsorption sites of the O atoms, the initial and final adsorption sites of the coadsorbed CO, the potential energy per simulation cell $E$ referred to that of the (lowest energy) structure A, and the height of the CO center of mass from the surface (defined as the average heights of the Ru atoms in the topmost layer) $Z_{\mathrm{CO-Ru}}$.}
    \label{tab:0875ml_adsorption}
\end{table}

Table~\ref{tab:0875ml_adsorption} summarizes the results of our structural search for the intermediate coverage.  In structure B, the CO molecules remain adsorbed in the hcp sites (Fig.~\ref{fig:0875ml_relaxation}, bottom panel). The lowest energy arrangement corresponds to structure A (Fig.~\ref{fig:0875ml_relaxation}, top panel). In this structure, which is the optimized structure when the CO molecules are initially on either top, fcc, or bridge sites, the CO molecules adsorb near the top site. Specifically, the CO molecules labeled as CO2 and CO3 in the figure are located in the line joining the top and bridge sites, with their axes slightly tilted towards the bridge site. The molecule labeled CO1 is located in the line joining the top and fcc sites and is slightly tilted towards the fcc site. Note that CO1 is situated between CO2 and the periodical image of CO3 (not shown in the figure). This means that the local coverage is higher around CO1 than around CO2 and CO3 that have an empty site in their vicinity. As a result, none of the O atoms are strictly equivalent and, though they remain very close to the hcp sites, they are slightly displaced along the $y$ direction (see Fig.~\ref{fig:0875ml_relaxation}). The values of these displacements respect to their initial hcp sites are $0.08$~{\AA}, $0.06$~{\AA}, $-0.124$~{\AA}, and $0.158$~{\AA} for O1, O2, O3, and O4, following this order. 

\subsubsection{CO desorption and oxidation on structure A}

The study of the reactivity is somewhat more complex than in the previous low coverage case due to the amount of non-equivalent CO and O adsorbates at the surface. 
After calculating all possible CO+O recombinations, we find however that there exist a number of energetically nearly equivalent recombinations due to the similar reorganization of the adsorbates that remain on the surface. In particular, using the labeling in Fig.~\ref{fig:0875ml_relaxation}, we obtain the following nearly equivalent recombinations: CO1+O1$\cong$CO1+O3, CO1+O2$\cong$CO1+O4, CO2+O1$\cong$CO3+O2, CO2+O2$\cong$CO3+O3, CO2+O3$\cong$CO3+O4, and CO2+O4$\cong$CO3+O1. The reactions energies for each CO and each CO+O recombination are given in Table~\ref{tab:0875ml_desorption}. Note that the reaction energies between the nearly equivalent recombinations differ in 1~meV at most.

\begin{table}
    \centering
    \begin{tabular}{ccc}
        \hline\hline
        Reaction  & Adsorption site         & $E$(eV) \\
        \hline
         CO$_\mathrm{(ads)} \rightharpoonup $ CO$_\mathrm{(gas)}$    & CO1         & 0.580 \\
        & CO2 (CO3)         & 0.730 (0.731) \\

                    &               &     \\
     O$_\mathrm{(ads)}$+CO$_\mathrm{(ads)} \rightharpoonup $ CO$_\mathrm{2(gas)}$ 
                   & CO1+O1 (CO1+O3)   & $-$0.169 ($-$0.169) \\
                    & CO1+O2 (CO1+O4)    & $-$0.678 ($-$0.679) \\
                    & CO2+O1 (CO3+O2)   & $-$0.815 ($-$0.814) \\
                    & CO2+O2 (CO3+O3)   & $-$0.176 ($-$0.176) \\
                    & CO2+O3 (CO3+O4)   & 0.026 (0.027)\\
                    & CO2+O4 (CO3+O1)& $-$0.584 ($-$0.584)\\
                    & &\\
        CO$_\mathrm{phys}\rightharpoonup $ CO$_\mathrm{(gas)}$    &  CO1$_\mathrm{phys}$         &  0.178 \\
        &  CO2$_\mathrm{phys}$ (CO3$_\mathrm{phys}$)         &  0.205 (0.201) \\
    \hline\hline
    \end{tabular}
    \caption{Reaction energies $E$ for CO desorption and oxidation from the Ru(0001) surface with 0.5~ML O + 0.375~ML CO coverage in  structure A (Fig.~\ref{fig:0875ml_relaxation}), calculated for all the possible reactants. The initial adsorption site for each desorbing species and each reaction is indicated in the second column following the labeling defined in Fig.~\ref{fig:0875ml_relaxation} for structure A. The desorption energies from the physisorption wells (CO1$_\mathrm{phys}$, CO2$_\mathrm{phys}$, CO3$_\mathrm{phys}$) are also provided. Note that the (energetically) quasiequivalent reactions and their energies are given within parenthesis.}
    \label{tab:0875ml_desorption}
\end{table}

As shown in Table~\ref{tab:0875ml_desorption}, CO desorption in structure A is less endothermic than in the honeycomb structure of the low coverage. In particular, the CO1 molecule, which experiences the highest local CO-coverage, has a desorption energy around 0.15~eV smaller than that of CO2/CO3 (0.58 vs 0.73~eV), and around 1~eV smaller than that of CO in the honeycomb structure. As a consequence, upon excitation of the surface, the desorption of CO is expected to be more efficient at this coverage than at the honeycomb low coverage. Remarkably, the oxidation of CO on structure A is exothermic for all possible recombinations, except for the pair (CO2+O3) (and its energetically quasiequivalent (CO3+O4)) that becomes endothermic by less than 30~meV (note that all the corresponding reaction energies in Table~\ref{tab:0875ml_desorption} are negative for the rest). Since CO1 is the CO adsorbate that requires less energy to be desorbed, one would expect that the energetically most favorable CO$_2$ desorption should also involve CO1. However, a counter intuitive result is obtained. It is the oxidation of CO2 with O1 (or, equivalently, CO3 with O2) the most exothermic reaction. We attribute this result to the different rearrangement of the adsorbates after CO oxidation. Our calculations for the two rearrangements show that the latter (i.e., without CO2+O1) is 0.136~eV more stable than the former (i.e., without CO1+O2). 

\begin{figure}
  \includegraphics[width=0.75\columnwidth]{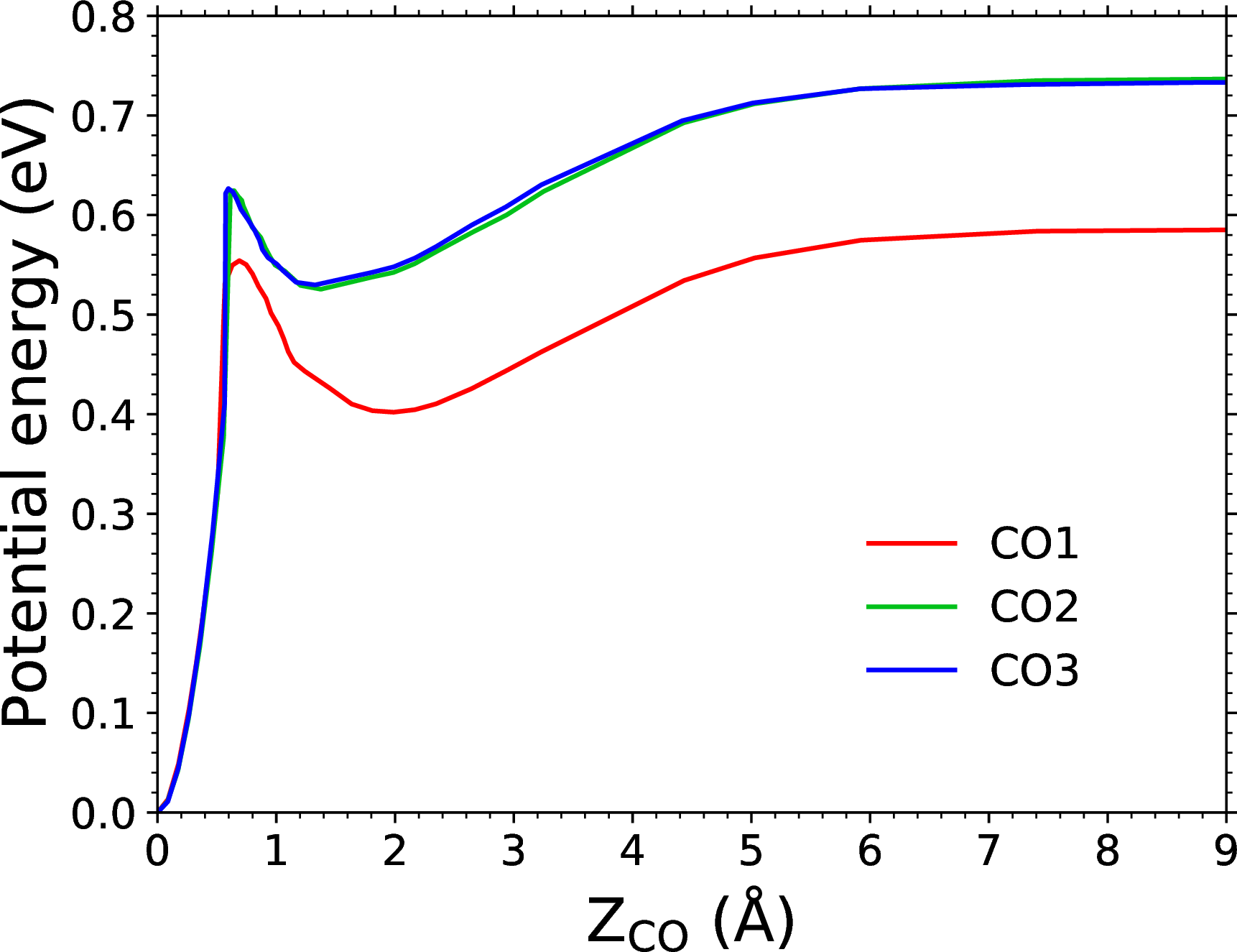}
\caption{Potential energy of a CO molecule as a function of its center of mass height $Z_{\mathrm{CO}}$, measured from its equilibrium adsorption position $Z_{\mathrm{CO-Ru}}$. Results obtained for the 0.5~ML O + 0.375~ML CO covered Ru(0001) surface in structure A (Fig.~\ref{fig:0875ml_relaxation}), for the CO1 molecule (red), CO2 molecule (green) and CO3 molecule (blue). The zero of energy is chosen for each species as that of its equilibrium adsorption position.}
  \label{fig:0875ml_co_desorption}
\end{figure}

Figure~\ref{fig:0875ml_co_desorption} shows the potential energy as a function of the distance from their adsorption sites for CO1, CO2, and CO3. 
The almost coincident CO2 and CO3 curves evidence that these adsorbates are nearly equivalent, as also observed when comparing their respective desorption and oxidation energies in Table~\ref{tab:0875ml_desorption}. The most important feature shown by the three curves is the existence of a physisorption well, absent in the case of low CO coverage. Note in passing that this well is not predicted by a GGA xc functional such as PBE, and thus, its existence remarks the importance of using a functional that incorporates the van der Waals interaction for this system. Common to the three curves, there is an incipient energy barrier separating the physisorption state from the chemisorption state. However, this barrier is in the three cases smaller than the corresponding energy barrier to desorption. In common to the low coverage, this result confirms that under the surface preparation followed in refs.~\citenum{Kostov1992Nov,Bonn1999Aug,Oberg2015Aug,Ostrom2015Feb} the 0.5~ML-O + 0.375-CO coverage is also achievable. Finally note that the asymptotic region is reached at slightly larger distances than in the low coverage case due to the existence of the physisorption well.

Our multiple attempts and strategies to obtain the MEP for the most exothermic oxidative reaction involving the pair CO2+O1 (or equivalently CO3+O2) have been totally unsuccessful. With none of our initial guess images we have been able to find stable intermediate states for this recombination (i.e., bCO$_2$ and lCO$_2$). We attribute this apparent lack of molecular adsorption states to the fact that CO2 is actually closely surrounded by the O2, O3, and O4 adsorbates. This situation makes difficult the approach of CO2 to O1 without altering substantially the (O2,O3,O4) arrangement. These considerations suggest the impossibility of this specific reaction or that the actual reaction path traverses through very high energy states. As an alternative, we have calculated the MEP for the energetically second most favorable recombination that involves the pair CO1+O2. The information and conclusions extracted from these results are meaningful, but we acknowledge that we cannot completely exclude the existence of a lower energy path involving the CO2 and O1 species. 

\begin{figure}
  \includegraphics[width=0.75\columnwidth]{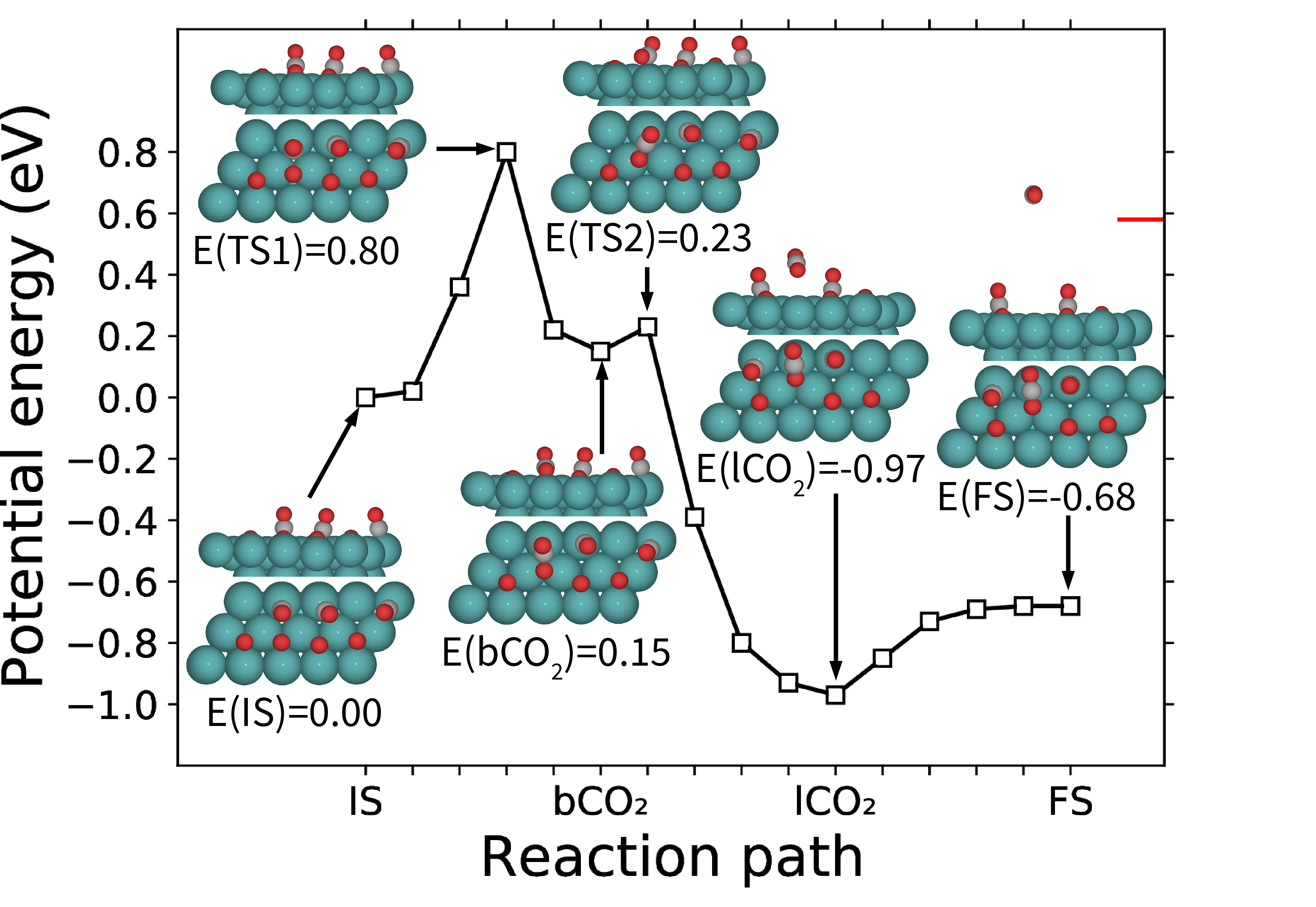}
 
 \caption{CI-NEB calculated minimum energy path for the recombinative desorption of CO1+O2 from structure A of the Ru(0001)-(0.5~ML O + 0.375~ML CO) surface. Also are shown the top and side views of the relevant configurations along the path: the initial state O$_\mathrm{(ads)}$+CO$_\mathrm{(ads)}$ (IS), the final state CO$_2\mathrm{(gas)}$ (FS), the intermediate adsorption states (bCO$_2$ and lCO$_2$), and the transition states (TS1 and TS2). Their energies referred to IS are given in eV below each image. For comparison, the energy of the desorbed CO is also given by a red short-line on the top right of the figure. Color code: O in red, C in gray, and Ru in blue.}
  \label{fig:0875ml_cineb}
\end{figure}

Figure~\ref{fig:0875ml_cineb} shows the calculated MEP for the recombinative desorption of CO1 and O2, including images of the main states. As in the low coverage, there are two minima along the MEP that correspond to the molecularly adsorbed states, bCO$_{2}$ and lCO$_{2}$, and two transition states, TS1 and TS2, that separate bCO$_{2}$ from IS and from lCO$_{2}$, respectively. The chemisorbed bCO$_{2}$ state with an energy of 0.15~eV respect to IS and with energy barriers to escape to lCO$_2$ and to IS of 0.08~eV and 0.65~eV, respectively, is also metastable in the intermediate coverage. The physisorbed state lCO$_{2}$, being $-0.97$~eV below IS and $-0.29$~eV below FS, becomes the lowest energy state in the reaction path at the intermediate coverage. The corresponding physisorption well is located around $1.5$~{\AA} above the equilibrium adsorption site of CO$_{\mathrm{(ads)}}$. Altogether, the activation barrier for CO oxidation is also governed at this coverage by TS1 and its value is 0.8~eV. In other words, though, as shown in Table~\ref{tab:0875ml_desorption}, the process is exothermic, it is necessary to give, at least, 0.8~eV to the system in order the CO oxidation can take place. In fact, this activation barrier for CO oxidation is larger by 0.22~eV than the energy required to desorb CO, suggesting that CO desorption would dominate over CO oxidation on the intermediate coverage. Nonetheless, the final answer would require to perform molecular dynamics simulations.

\subsection{High coverage:  0.5~ML O + 0.5~ML CO}

The proposed (O,CO)-mixed monolayer over the Ru(0001) surface, which is simulated by four adsorbed O and four adsorbed CO molecules in our 4$\times$2 simulation cell, has not been obtained experimentally yet. However, as we will show below, it is a stable structure. 

\begin{figure}
  \includegraphics[width=0.75\columnwidth]{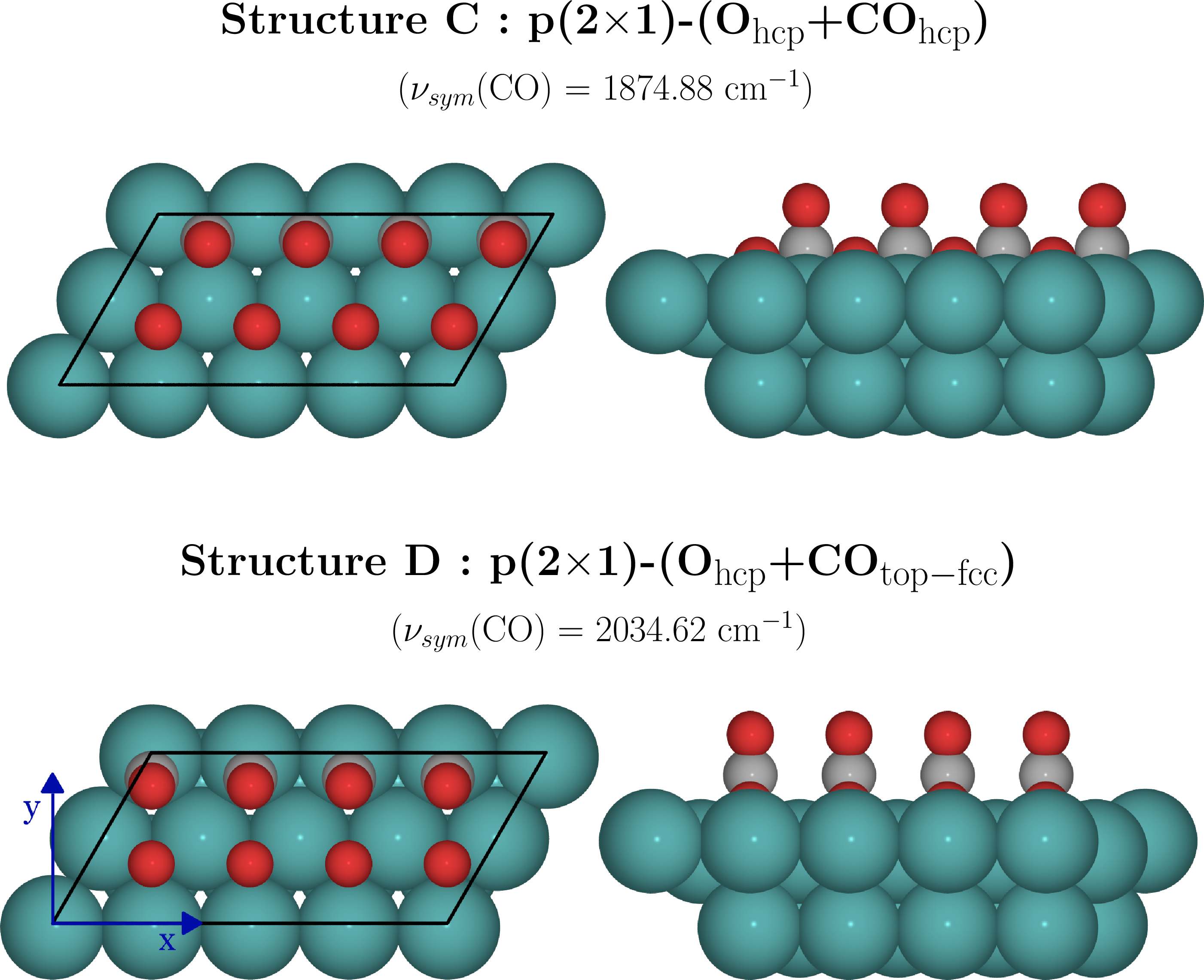}
\caption{Top (left) and side (right) views of the two stable structures obtained for the 0.5~ML O + 0.5~ML CO coverage on Ru(0001). Top panels: the energetically most stable  p(2$\times$1)-(O$_\mathrm{hcp}$+CO$_\mathrm{hcp}$) structure. Bottom panel: p(2$\times$1)-(O$_\mathrm{hcp}$+CO$_\mathrm{top-fcc}$) structure.
The frequency of the in-phase C--O stretch mode is specified for each coverage. Color code: O atoms in red, C in gray, and Ru in blue. The black parallelograms show the surface unit cells in the calculations. For clarity, the periodic images of the O and CO adsorbates are not shown and only the two topmost Ru layers are depicted in the side views.}  \label{fig:1ml_relaxation}
\end{figure}

We start studying the stability of those structures that are compatible with the initial p(2$\times$1)-O overlayer that is formed in the experiments prior dosing the surface with CO. Thus, coadsorbed with p(2$\times$1)-O$_\mathrm{hcp}$, in which the four O atoms in our 4$\times$2 unit cell adsorb in hcp sites, we consider the six possible p(2$\times$1)-CO arrangements depicted in Fig.~S4 of the Supporting Information. In five of these guess structures, all the CO molecules adsorb on equivalent sites, that is, on either hcp, fcc, top, bridge-a, or bridge-c sites. In the sixth structure, denoted bridge-b, the CO molecules are equally distributed among the two nonequivalent bridge sites. Upon relaxation the adsorbates stabilize into one of the two structures represented in Fig.~\ref{fig:1ml_relaxation}. The only exception is the guess structure with the CO molecules initially adsorbed in on-top sites that end desorbing. 
The p(2$\times$1)-(O$_\mathrm{hcp}$+CO$_\mathrm{hcp}$) structure is obtained when the CO adsorbates are initially on either hcp,  bridge-b, or bridge-c. In this structure (hereafter denoted structure C), the CO molecules are located in the hcp sites with the molecular axis slightly tilted. 
This is the lowest energy structure at the high coverage. The second stable arrangement, p(2$\times$1)-(O$_\mathrm{hcp}$+CO$_\mathrm{top-fcc}$) (hereafter denoted structure D),
is found when the CO molecules are initially on either fcc or bridge-a sites. In this case, CO locates in the line joining the top and fcc sites with its molecular axis slightly tilted toward the fcc site. The energy of structure D is around $0.29$~eV per simulation cell higher than that of the equilibrium structure C. The results of the structural search for the high coverage are summarized in Table~\ref{tab:1ml_adsorption}. As previously, the distinct frequencies of the in-phase C--O stretch mode are given in the figure for the two optimized structures (Fig.~\ref{fig:1ml_relaxation}). In common to the low and intermediate coverages, the highest frequency corresponds to CO adsorbed on near-top position. Interestingly, the high-coverage frequencies are around 30~cm\textsuperscript{-1} larger than the ones found for the intermediate and low coverages at similar adsorption sites. We ascribe this effect to the weaker CO-surface bound formed in the high coverage, that is reflected in the smaller CO adsorption energies that we obtain for this coverage (see below) as compared to the others. This effect, i.e., the increase of the C--O stretch frequency with CO coverage (for the same adsorption site) has been observed and discussed for different metal surfaces~\cite{loffreda99, shanjpcc09,gunasooriya}. 

\begin{table}
    \centering
    \begin{tabular}{cccccc}
        \hline\hline
        O-structure & O site & CO initial site & CO final site & $E$(eV) & Z$_{\mathrm{CO-Ru}}$({\AA})\\
        \hline
        p(2$\times$1)   & hcp   & top           & desorbed      & -- & -- \\
        p(2$\times$1)   & hcp   & hcp           & hcp       & 0.000 & 2.20 \\
        p(2$\times$1)   & hcp     & bridge-b    & hcp       & 0.000 & 2.20 \\
        p(2$\times$1)   & hcp     & bridge-c    & hcp       & 0.000 & 2.20 \\
        p(2$\times$1)  &  hcp     & fcc           & top-fcc  & 0.287 & 2.63 \\
        p(2$\times$1)  &  hcp     & bridge-a    & top-fcc  & 0.292 & 2.63 \\
        \hline\hline
    \end{tabular}
\caption{Results from the structural optimization for the 0.5~ML O + 0.5~ML CO coverage on Ru(0001), indicating the structure and adsorption sites of the O atoms, the initial and final adsorption sites of the coadsorbed p(2$\times$1)-CO, the potential energy per simulation cell $E$ referred to that of structure C, which is the lowest energy arrangement, and the height of the CO center of mass from the surface (defined as the average heights of the Ru atoms in the topmost layer) $Z_{\mathrm{CO-Ru}}$.}
    \label{tab:1ml_adsorption}
\end{table}

\begin{figure}
  \includegraphics[width=0.75\columnwidth]{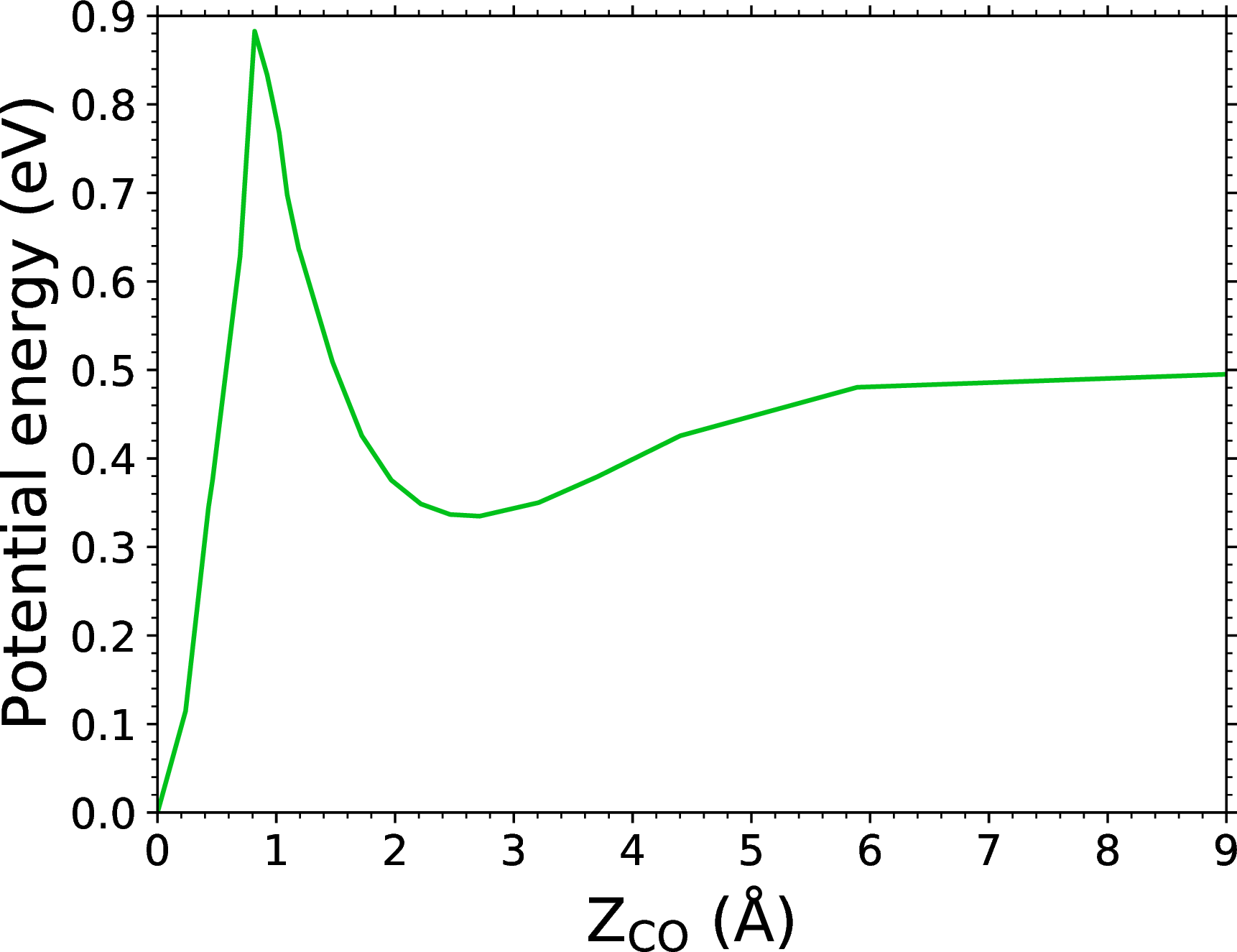}
 \caption{Potential energy of a CO molecule as a function of its center of mass height $Z_{\mathrm{CO}}$, measured from its equilibrium adsorption position $Z_{\mathrm{CO-Ru}}$. Results obtained for the 0.5~ML O + 0.5~ML CO coverage in the optimized structure C.
 The zero of energy is chosen as that of the equilibrium adsorption position.}
  \label{fig:1ml_co_desorption}
\end{figure}

\subsubsection{CO desorption and oxidation on structure C}

Table~\ref{tab:1ml_desorption} shows the calculated reaction energies for CO desorption and oxidation on structure C of high coverage. Since at this coverage all the CO molecules and all the O adsorbates are equivalent, the oxidation energy is only calculated for a pair of nearest CO and O neighbors. In common to structure A of the intermediate coverage, the energy values show that oxidation of CO is exothermic and the desorption endothermic. At the high coverage, the difference between the two reactions energies is around $1.88$~eV, which is a notable value. However, as in previous cases, energy barriers could exist in the reaction paths that would imply the existence of activation energies governing the process even in the exothermic cases. Note that the endothermicity of CO desorption implies the stability of this structure that we have further confirmed by means of ab initio atomistic thermodynamics~\cite{Reuter2001,Zhao2015Jul} (see section S7 in Supporting Information).

\begin{table}
    \centering
    \begin{tabular}{ccr}
        \hline\hline
        Reaction  & adsorption site         & $E$(eV) \\
        \hline
        CO$_\mathrm{(ads)} \rightharpoonup$ CO$_\mathrm{(gas)}$          & CO$_\mathrm{hcp}$           & 0.488 \\
        O$_\mathrm{(ads)}$+CO$_\mathrm{(ads)} \rightharpoonup $ CO$_\mathrm{2(gas)}$        & CO$_\mathrm{hcp}$, O$_\mathrm{hcp}$    & $-$1.389 \\
        CO$_\mathrm{(phys)} \rightharpoonup$ CO$_\mathrm{(gas)}$          & CO$_\mathrm{phys}$           & 0.155 \\
    \hline\hline
    \end{tabular}
     \caption{Reaction energies $E$ for CO desorption and oxidation from the Ru(0001) surface with 0.5~ML O + 0.5~ML CO coverage in structure C. The initial adsorption site for each desorbing species is indicated as a subscript. The desorption energy from the physisorption well (CO$_\mathrm{phys}$) is also provided.}
    \label{tab:1ml_desorption}
\end{table}

Figure~\ref{fig:1ml_co_desorption} shows the value of the potential energy as one CO desorbs from structure C of high coverage. Since at this coverage all the CO adsorbates are equivalent, only a single curve is calculated. Alike the intermediate coverage, the CO desorption curve confirms the existence of a physisorption well that in this case is located at around
2.7~{\AA} from the chemisorption well. There is a new feature appearing in the high coverage that was not present at the lower coverages. The energy barrier of $0.883$~eV separating both adsorption wells is notably larger than the calculated chemisorption energy of $0.488$~eV and hence rules the desorption process. Additionally, the existence of this barrier can explain why this high coverage has not been achieved with the usual experimental techniques based on thermal deposition of CO under UHV conditions on the previously O-saturated surface. Indeed, it is extremely unlikely that thermally deposited CO molecules could gain the $0.395$~eV required to overcome the barrier when coming from vacuum. In other words, the presence of this barrier is expected to prevent the CO molecules from being chemisorbed, so that this high coverage could be achieved.

\begin{figure}
  \includegraphics[width=0.75\columnwidth]{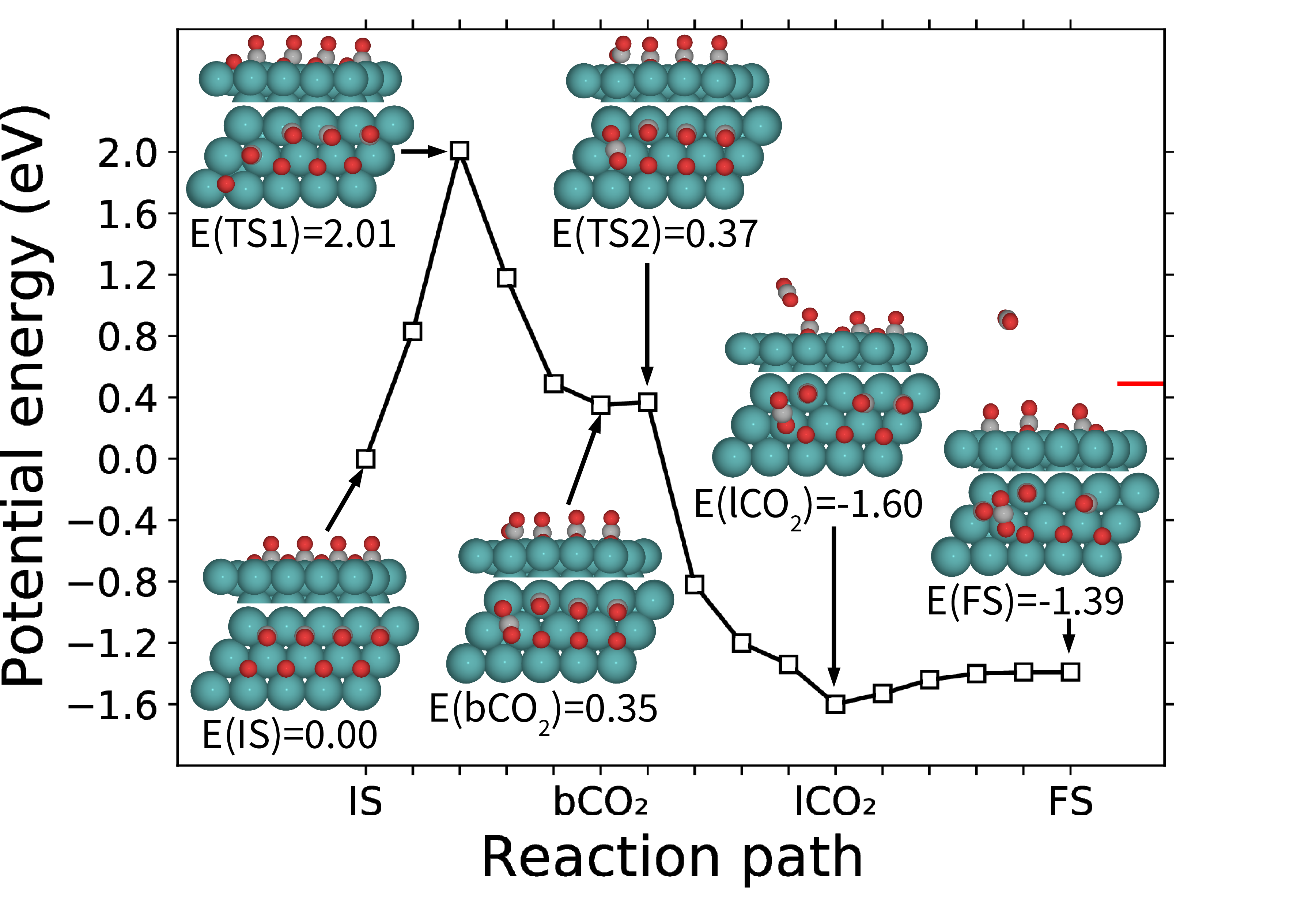}
   \caption{CI-NEB calculated minimum energy path (MEP) for the CO$_2$ recombinative desorption from Ru(0001) with the p(2$\times$1)-(O$_\mathrm{hcp}$+CO$_\mathrm{hcp}$) coverage. Also are shown the top and side views of the relevant configurations along the path: the initial state O$_\mathrm{(ads)}$+CO$_\mathrm{(ads)}$ (IS), the final state CO$_2\mathrm{(gas)}$ (FS), the intermediate adsorption states (bCO$_2$ and lCO$_2$), and the transition states (TS1 and TS2). Their energies referred to IS are given in eV below each image. For comparison, the energy of the desorbed CO is also given by a red short-line on the top right of the figure. Color code: O in red, C in gray, and Ru in blue.}
  \label{fig:1ml_cineb}
\end{figure}

The calculated MEP for the oxidation of CO is shown in Fig.~\ref{fig:1ml_cineb} together with images of the configurations that correspond to the extremes of the path. As in the low and intermediate coverages, we identify two molecular adsorption states, the chemisorbed bCO$_{2}$ and the physisorbed lCO$_2$, and the corresponding transition states TS1 and TS2, separating the chemisorbed state from IS and from the physisorbed state, respectively. Qualitatively, the MEP is similar to the one found at the intermediate coverage. In particular, lCO$_2$ is the lowest energy state in the reaction path with an energy of $-1.60$ ($-0.21$)~eV respect to IS (FS), while TS1 is the highest energy state located $2.01$~eV above IS \footnote{In order to converge the subpath from IS to bCO$_2$ it was necessary to restrict the degrees of freedom to those of the recombining O+CO, while keeping the rest fixed. The ulterior relaxation of the TS1 configuration reduced the energy to 1.72~eV, but the new configuration is not a saddle point. Therefore, we estimate that the real TS1 must be in the range 1.72-2.01~eV.}.
As in previous coverages, bCO$_2$ is metastable, being $1.66$~eV and $0.02$~eV the energy barriers to escape from this state to the dissociatively adsorbed O$_\mathrm{(ads)}$+CO$_\mathrm{(ads)}$ state and to lCO$_2$, respectively. Regarding the comparison with the CO desorption process, we observe that though, as stated previously, CO desorption is endothermic and CO$_2$ desorption exothermic, the activation barrier for this last process is not only much higher than the energy of the desorbed CO state ($0.488$~eV), but also much higher than the activation energy for CO desorption ($0.883$~eV). 

\subsection{Coverage Comparison}

In comparing the coverage dependence of the CO desorption and oxidation, we consider meaningful to include in the discussion the results obtained in the ideal (non-interacting) O+CO zero coverage limit that allows to describe the oxidation process (denoted O+CO-Ru(0001) in the following). This is the case of having a single O and a single CO separately adsorbed on the Ru(0001) surface. Under these conditions, the molecule adsorbs atop one Ru atom (similarly to the aforesaid CO zero coverage limit) and O in a distant hcp site (see the Supporting Information for further details). The desorption and oxidation of CO on this surface are both highly endothermic, being the energies very similar, 1.59 and 1.63~eV, respectively. As in the previous coverages, the oxidation MEP is characterized by two (local) minima that correspond to the chemisorbed bCO$_{2}$ and physisorbed lCO$_{2}$ states and by the corresponding maxima TS1 and TS2. It is meaningful to compare these results to those obtained by Zhao et al.~\cite{Zhao2017Jun} also in the, essentially, zero coverage limit. In that work they used the RPBE-GGA exchange-correlation functional~\cite{hammer99}, which neglects the van der Waals interaction, and a 5$\times$5 simulation cell. The results obtained in our case for the energy of the TS1 and bCO$_{2}$ states measured from IS are, considering the different calculation setups, in rather good agreement with those obtained by Zhao et al., with energy differences that amount less than 60~meV. However, their CO$_2$ desorption energy is 0.48~eV lower than ours, probably due to the too repulsive RPBE functional. Moreover, the lCO$_{2}$ and TS2 states were not reported by them. This should be attributed to the neglect of the van der Waals interaction, which, as discussed above, is necessary to describe the physisorption region.

\begin{figure}
  \includegraphics[width=0.75\columnwidth]{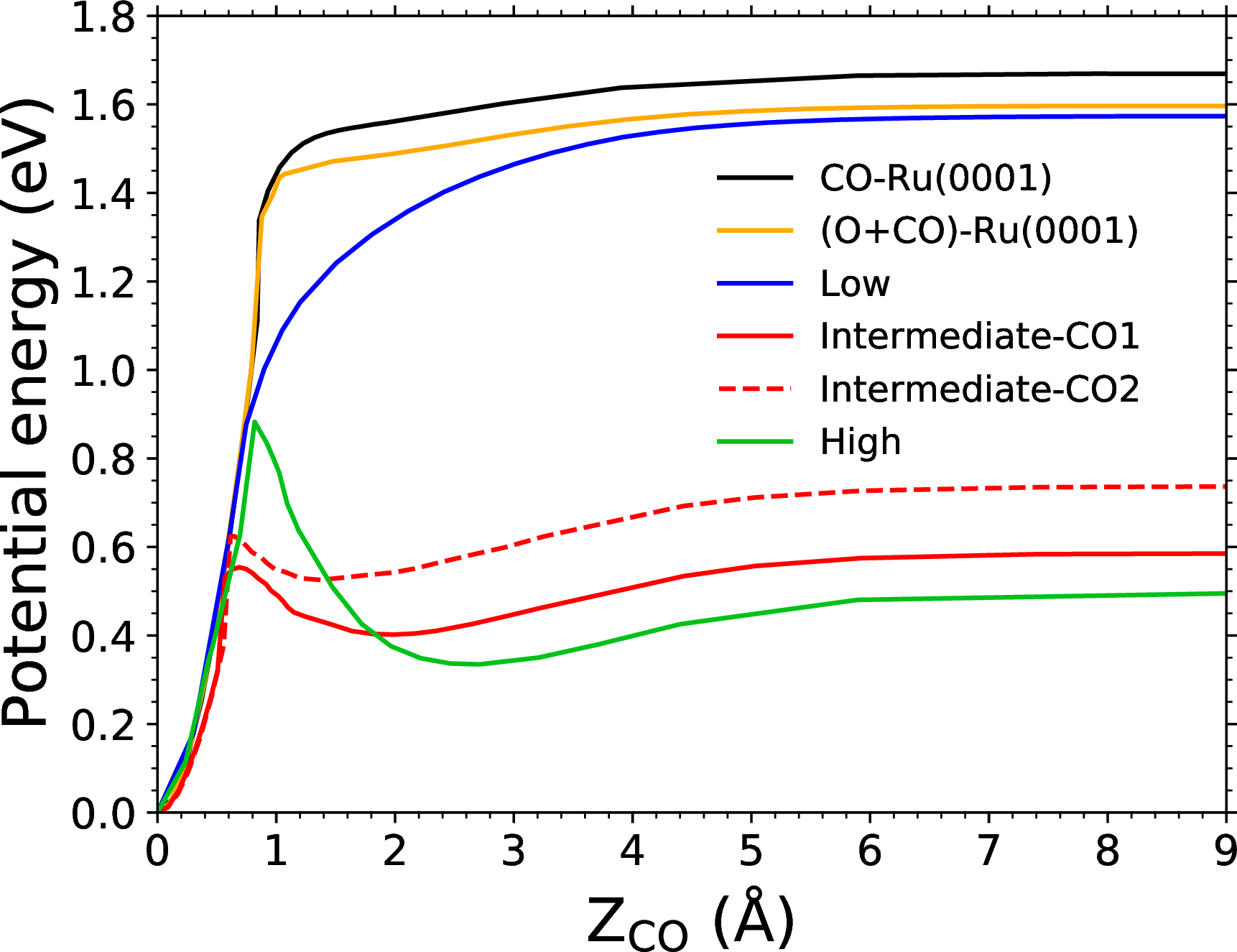}
  \caption{Potential energy of a CO molecule as a function of its center of mass height $Z_{\mathrm{CO}}$, measured from its equilibrium adsorption position $Z_{\mathrm{CO-Ru}}$ at different coverages in their optimized structure. In all cases, the zero of energy is chosen as that of the corresponding equilibrium adsorption position. In the case of the intermediate coverage results for the nonequivalent CO1 (solid line) and CO2 (dashed line) adsorbates are shown.}
  \label{fig:all_co_desorption}
\end{figure}

The CO desorption curves are compared in Fig.~\ref{fig:all_co_desorption} for the different coverages. As a general trend, we observe a decrease in the desorption energy as the coverage increases. Desorption from the CO-Ru(0001) and O+CO-Ru(0001) surfaces is very similar, except for a minor reduction of around $0.068$~eV in the desorption energy in the presence of O. Increasing the O and CO coverages to 0.5~ML and 0.25~ML causes a further reduction of  $0.02$~eV. Clearly, the overall CO desorption energetics is very similar in these three low coverages, as an indication of the minor influence exerted by the O adsorbates. In contrast, the CO desorption energy is drastically reduced when increasing the CO coverage from 0.25~ML to 0.375 and 0.5~ML, showing the limit at which the CO-CO dipole interaction becomes relevant and competes against the CO-surface binding. This is a common effect widely observed by different authors on different metals~\cite{alducinprl19, loffreda99, steckelprb03, shanjpcc09, martinjpcc14, gunasooriya}. Also the appearance of a stable physisorption well at these two large CO-coverages seems to be a consequence of the CO-CO interaction. The depth of the wells is similar for both coverages ($E_\mathrm{phys}$(CO1)=$0.178$~eV and $E_\mathrm{phys}$(CO2)=$0.205$~eV at the intermediate coverage and $E_\mathrm{phys}$=$0.155$~eV at the high coverage, well depths measured from vacuum), but there are also differences between them. The distance between the chemisorption and physisorption wells is $0.7$~{\AA} smaller in the case of intermediate-CO1 than at high coverage, and $1.3$~{\AA} smaller in the case of intermediate-CO2 than at high coverage. As a result, the energy barrier separating the molecularly chemisorbed and physisorbed states is visibly higher at the high coverage than at the intermediate coverage. A simplistic explanation can be obtained in terms of a 1D picture of the Lennard-Jones model for chemisorption and physisorption at surfaces. According to this simplified model,  the barrier will appear at the crossing between the chemisorption and physisorption curves and therefore, it will increase as the separation between both wells increases.  In both coverages (intermediate and high), the existence of the physisorption well and concomitant energy barrier is expected to limit the access to the chemisorbed state from gas-phase. In particular, completion of the 0.5 ML CO coverage is prevented by a large positive barrier of $0.395$~eV and, as aforementioned, it explains that this coverage is not formed in the experiments by Kostov et al.~\cite{Kostov1992Nov}. 

\begin{figure}
  \includegraphics[width=0.75\columnwidth]{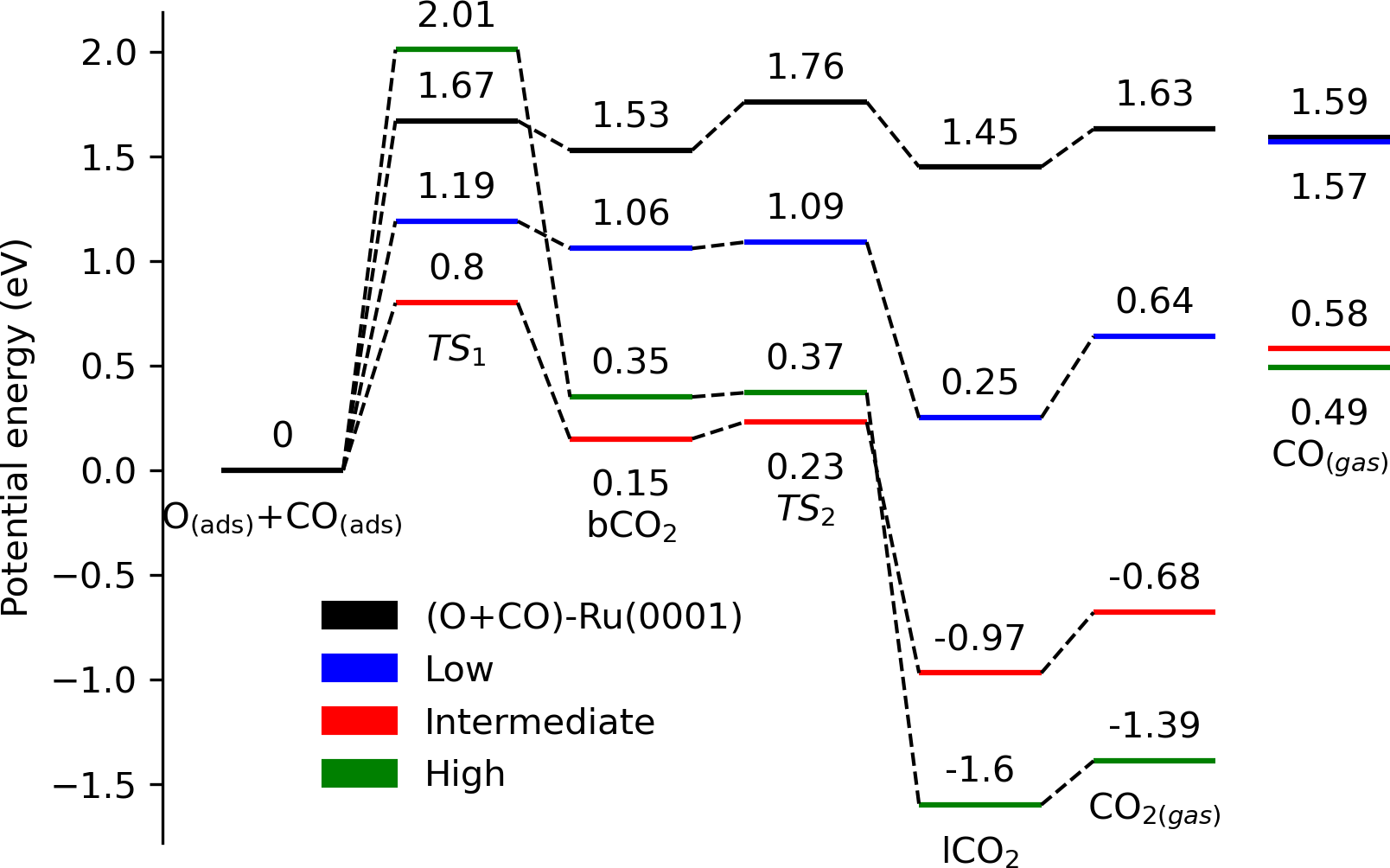}
\caption{Schematic energy diagram comparing the oxidation of CO on the different covered surfaces for the relevant states along the MEP, including the initial (IS) and final (FS) states, as well as the extremes of the paths that are defined by the two molecular adsorption states (bCO$_2$ and lCO$_2$) and the two transition states (TS1 and TS2). For comparison the CO desorption energy is also shown for all coverages.}
  \label{fig:all_cineb}
\end{figure}

In the diagram of Fig.~\ref{fig:all_cineb} we compare the calculated MEP for the oxidation of CO on the different coverages, including the ideal zero coverage limit O+CO-Ru(0001). The figure shows schematically the energies of the relevant states along each reaction path, namely, the initial (IS) and final (FS) states, the two molecular adsorption states (bCO$_2$ and lCO$_2$), and the two transition states (TS1 and TS2). As observed for the CO desorption energy, the oxidation energy decreases as the coverage increases, although exhibiting a more monotonic dependence. As a result, the reaction changes from endothermic at the zero and low coverages to exothermic at the intermediate and high coverages. A similar behavior is observed for oxidation into the lCO$_2$ state.  However, the results for bCO$_2$, TS1, and TS2 at the high coverage break this tendency, since their corresponding energies are higher at the high coverage than at the intermediate coverage. This sudden increase can be due to the difficulty of forming the bCO$_{2}$ on the overcrowded surface, where the near adsorbates are necessarily too close and contribute to the energy increase. As a final observation, note that TS1 is the transition state with the highest energy and, therefore, governs the reaction at the low, intermediate, and high coverages. For the O+CO zero coverage limit, TS2 is $0.09$~eV more energetic than TS1 and constitutes the limiting step for CO oxidation in this case.

In principle, differences in the charge state of the adsorbates may contribute to explain that the oxidation energetics changes from one coverage to another. To this aim, we have calculated the charge state along the minima of the MEP for each coverage. Interestingly, there is no apparent dependence on the coverage to relate to the different barriers and adsorption energies. This is shown in Table~\ref{tab:charge}, where we reproduce the charge state $Q$ of the recombining CO$_2$ at relevant states along the oxidation MEP for all the coverages so far studied. The charge state of each atom forming the CO$_2$ molecule, as well as the charge state of those forming the adsorbed CO and of the adsorbed O are provided in the Supporting Information. Starting with IS, there is a noticeable electron transfer from the surface to the O and CO adsorbates. The atom-resolved analysis provided in the Supporting Information shows that the electronegative O captures about 0.7$e^-$ and  roughly 0.1$e^-$. Formation of bCO$_2$ causes a small reduction in $Q$ for all coverages. Still the molecule retains around 0.6 electrons from the surface that evidences the chemisorbed nature of the bCO$_2$ state in all the cases.
In the lCO$_2$ state, the charge state becomes zero as a clear indication of the physisorbed character of the bonding. Finally, note that the charge state of the desorbed CO$_{2(gas)}$ is zero as it should. 
\begin{table}
    \centering
    \begin{tabular}{ccccc}
        \hline\hline
        coverage  & IS & bCO$_2$ & lCO$_2$ &  FS \\
        \hline
        (O+CO)-Ru(0001)  & $-$1.067 &$-$0.586&$-$0.040&0.0 \\
        0.5ML O+0.25ML CO  & $-$0.856 &$-$0.595&$-$0.003&0.0 \\
        0.5ML O+0.375 ML CO& $-$0.803 &$-$0.581&$-$0.012&0.0 \\
        0.5ML O+0.5 ML CO & $-$0.889 &$-$0.515&0.0&0.0\\
    \hline\hline
    \end{tabular}

     \caption{Charge state $Q$ (see eq.\eqref{eq:Q}) of the recombined CO$_2$ at different configurations along the oxidative MEP found for each optimized coverage: the initial O$_\mathrm{(ads)}$+CO$_\mathrm{(ads)}$ state (IS), the chemisorbed bCO$_2$ state, the physisorbed lCO$_2$ state, and the final CO$_\mathrm{2(gas)}$ state (FS).}
    \label{tab:charge}
\end{table}

\section{Summary}

The coadsorption of O and CO on covered Ru(0001) surfaces is studied with DFT and the exchange-correlation functional of Dion et al.~\cite{Dion2004Jun} that includes non-local correlation corrections. Three coverages are considered that are compatible with the preparation of the mixed overlayer in experiments aimed to understand the competition between CO desorption and oxidation and the dependence on coverage~\cite{Kostov1992Nov,Oberg2015Aug,Ostrom2015Feb,Bonn1999Aug}. The experimental procedure consists in saturating the Ru(0001) surface with oxygen up to complete a 0.5~ML and next dose the precovered surface with CO. Therefore, coadsorbed with the oxygen saturation coverage of 0.5~ML we examine the following coverages: (i) the low coverage of 0.5ML O+0.25ML CO, (ii) the intermediate coverage of 0.5ML O+0.375ML CO, and (iii) the high coverage of 0.5ML O+0.5ML CO. Firstly, we identify the optimized lowest energy structure for each coverage. At the low coverage, the O and CO adsorbates arrange on the honeycomb structure reported experimentally, in which the CO adsorbed on top of the Ru atom surrounded by the honeycomb arrangement of the adsorbed O. At the intermediate coverage, the CO molecules adsorb in near-top sites along the empty rows left by the O atoms. The latter forms a p(2$\times$1) structure occupying the hcp sites. At the high coverage, p(2$\times$1)-O$_\mathrm{hcp}$ intercalates with a p(2$\times$1)-CO$_\mathrm{hcp}$ arrangement. Next, the energetics of CO desorption and oxidation is analysed for each of these lowest energy structures, as well as for the CO and O+CO zero coverage limit.

CO desorption is endothermic at all considered coverages, although the energy decreases as the coverage increases. The desorption path at the low coverage as well as at the CO and O+CO zero coverage limits is rather direct and simple since it only involves the chemisorption well. At the intermediate and high coverages however, new physisorbed CO states appear that may contribute to slow down the desorption process. Remarkably, at the high coverage the energy barrier separating both molecularly adsorbed states exceeds the energy of the molecule in vacuum by $0.395$~eV. It is precisely the presence of this barrier that prevents the gas-phase CO from being chemisorbed up to complete the 0.5~ML of CO on the precovered Ru(0001)-0.5~ML-O surface. 

The recombinative desorption of adsorbed O and CO, that is, the oxidation of CO on the surface is assumed to proceed trough the two molecularly adsorbed states that exist in all the coverages, the chemisorbed bCO$_{2}$ and the physisorbed lCO$_{2}$. The initial O$_\mathrm{(ads)}$+CO$_\mathrm{(ads)}$ recombination into bCO$_{2}$ is for all coverages endothermic. Also the recombination into lCO$_{2}$ is endothermic for both the low coverage and the O+CO zero coverage limit, but becomes exothermic for the intermediate and high coverages. Interestingly, this observation together with the appearance of the physisorbed CO state at the same coverages, suggest the importance of the dipolar interaction with the CO adsorbates in stabilizing the physisorbed states. The endothermiticity of the chemisorbed bCO$_2$ also decreases as the CO coverage increases, except for the high coverage. In this case, the energy of the bCO$_2$ state (respect to the O$_\mathrm{(ads)}$+CO$_\mathrm{(ads)}$ state) is about 0.2~eV larger than in the intermediate coverage. We ascribe this increase to the difficulty of forming CO$_2$ in presence of multiple adsorbates. Using CI-NEB between each adsorbed state we find two transition states for each of the four mixed coverages. The oxidation process is limited by the initial transition state that gives access to the chemisorbed bCO$_{2}$ in all the coverages, excluding the (O,CO) zero coverage limit. In this case, the process is ruled by the transition state between bCO$_{2}$ and lCO$_{2}$. 

Regarding the competition between CO desorption and oxidation, for all cases in which 0.5 ML O is adsorbed at the surface the CO desorption reaction is more endothermic than the CO$_2$ recombinative desorption. Among the studied coverages the only exception to this rule is the O+CO zero coverage limit, for which CO oxidation is $0.04$~eV more energetic than CO desorption. Nevertheless, except in the case of the low coverage honeycomb structure, for the rest of analysed coverages there exists at least one intermediate state along the CO oxidation path with energy higher than the CO desorption energy. In the high coverage case, the only case in which CO desorption is also governed by an energy barrier in the reaction path, its energy is more than $1$~eV lower than the activation energy for CO$_2$ recombinative desorption. Though this study of the energetics already constitutes an explanation for the prevalence of CO desorption over oxidation in experiments, other factors are expected to contribute and further reduce the relative likeness of the latter reaction. In order to desorb CO$_2$, a strongly bound O adsorbate must be destabilized, next, the adsorbed CO and O must diffuse on the surface and, finally, they must encounter under proper geometrical and energy conditions in order the molecular bond can be formed. This means that the configurational space for CO oxidation is expected to be much more limited than that for the more direct CO desorption with the subsequent reduction of the reaction probability. For this reason, even for the honeycomb structure in which CO$_2$ desorption is energetically more favorable, the reduced configuration space for oxidation can explain the prevalence of CO desorption over oxidation that is observed in experiments at this coverage.
The final answer to this and other questions will require a dynamics study that will be the focus of our future work.


\section*{Authors information}

\textbf{A. Tetenoire} --
Donostia International Physics Center (DIPC),
Paseo Manuel de Lardizabal 4, 20018 Donostia-San Sebasti\'an, Spain; orcid: https://orcid.org/0000-0001-7538-7543; email: auguste.tetenoire@dipc.org

\noindent \textbf{J.I. Juaristi} -- Departamento de Pol\'{\i}meros y Materiales Avanzados: F\'{\i}sica, Qu\'{\i}mica y Tecnolog\'{\i}a, Facultad de Qu\'{\i}micas (UPV/EHU), Apartado 1072, 20080 Donostia-San Sebasti\'an, Spain\\
Centro de F\'{\i}sica de Materiales CFM/MPC (CSIC-UPV/EHU), Paseo Manuel de Lardizabal 5, 20018 Donostia-San Sebasti\'an, Spain\\ Donostia International Physics Center (DIPC),
Paseo Manuel de Lardizabal 4, 20018 Donostia-San Sebasti\'an, Spain; orcid: https://orcid.org/0000-0002-4208-8464; email: josebainaki.juaristi@ehu.eus

\noindent \textbf{M. Alducin} -- Centro de F\'{\i}sica de Materiales CFM/MPC (CSIC-UPV/EHU), Paseo Manuel de Lardizabal 5, 20018 Donostia-San Sebasti\'an, Spain\\ Donostia International Physics Center (DIPC),
Paseo Manuel de Lardizabal 4, 20018 Donostia-San Sebasti\'an, Spain; orcid: https://orcid.org/0000-0002-1264-7034; email: maite.alducin@ehu.eus

\subsection*{Authorship contribution statement}

All authors contribute to the development of this project.

\subsection*{Declaration of Competing Interest}
Conflicts of Interest: None.

\begin{acknowledgement}
The authors acknowledge financial support by the Gobierno Vasco-UPV/EHU
Project No. IT1246-19 and the Spanish Ministerio de Ciencia e Innovaci\'on [Grant No. PID2019-107396GB-I00/AEI/10.13039/501100011033]. We thank Dr. Oihana Galparsoro for fruitful discussions during the final preparation of the manuscript. This research was conducted in the scope of the Transnational Common Laboratory (LTC) “QuantumChemPhys – Theoretical Chemistry and Physics at the Quantum Scale”. Computational resources were provided by the DIPC computing center.
\end{acknowledgement}

\begin{suppinfo}
The Supporting Information is available free of charge at XXX and includes further information on:
\begin{itemize}
    \item Selection of the exchange-correlation functional based on the CO adsorption energy
    \item Details of the structural search performed for each coverage
    \item Details of the minimum energy path for CO oxidation on the different covered surfaces under consideration: intermediate adsorption states, charge state analysis, geometries of the configurations along each MEP. 
\end{itemize}
\end{suppinfo}


\bibliography{general_biblio}

\end{document}